\documentclass[preprint,aps,amsmath,nofootinbib]{revtex4-1}
\usepackage{graphicx,array,dcolumn}
\usepackage{calc,tabularx, epsfig,mathrsfs}
\usepackage{hyperref}

\usepackage{amsmath,verbatim,enumerate}
\usepackage{amssymb}
\usepackage{wasysym}
\usepackage{tikz}
\usepackage{xcolor}
\usepackage{subfigure}
\usepackage{subcaption}
\usepackage[percent]{overpic} 
\usepackage{multirow}
\usepackage{xspace}
\usepackage{mathrsfs}
\usepackage{slashed}
\usepackage{mathtools}
\usepackage{cancel}
\usepackage[justification=raggedright,singlelinecheck=off]{caption}
\usepackage{pgfplots}
\pgfplotsset{compat=1.18}
\allowdisplaybreaks[1]
\newlength{\figurewidth}
\newcommand{\beq}{\begin{equation}}
\newcommand{\eeq}{\end{equation}}
\newcommand{\bea}{\begin{eqnarray}}
\newcommand{\eea}{\end{eqnarray}}
\newcommand{\ba}{\begin{array}}
\newcommand{\ea}{\end{array}}

\newcommand{\ep}{\epsilon}

\newcommand{\lam}{\lambda}

\makeatother
\begin{document}
%
%%%%%%%%%%%%%%%%%%%%%%%%%%%%%%%%%%%%%%%%%%%%%%%
\title{
%Axionic wormholes in Lorentzian path integral, \\
Boundary conditions for axionic wormholes, \\
imaginary distance bound and KSW allowability
%Boundary conditions for Euclidean wormholes, \\
%Axionic wormholes as complex saddles, \\
}
\setlength{\figurewidth}{\columnwidth}
%%%%%%%%%%%%%%%%%%%%%%%%%%%%%%%%%%%%%%%%%%%%%%%
%

\author{Shubhashis Mallik} 
\email{shubhashism@iisc.ac.in}

\author{Neha}
\email{neha2024@iisc.ac.in}

\author{Gaurav Narain}
\email{gnarain@iisc.ac.in}

\affiliation{
Center for High Energy Physics, Indian Institute of Science,
C V Raman Road, Bangalore 560012, India.
}

%%%%%%%%%%%%%%%%%%%%%%%%%%%%%%%%%%%%%%%%%%%%%%%
\vspace{50mm}
%%%%%%%%%%%%%%%%%%%%%%%%%%%%%%%%%%%%%%%%%%%%%%%

\begin{abstract}
We study four-dimensional axion wormholes in the Lorentzian minisuperspace path integral using dual scalar and three-form flux formulations. We analyze this duality for generic metric boundary conditions and real lapse integration contour, and show that it holds on any background, including complex ones. With the Dirichlet condition on the metric, the fixed-flux path integral is evaluated exactly. In the Euclidean regime, saddle geometries organize into same-side and cross-throat segments of the Giddings–Strominger (GS) wormhole, depending on whether the two boundaries lie on the same or opposite sides of the throat. A Picard–Lefschetz analysis in the covering plane of lapse reveals that the cross-throat saddle has vanishing intersection number, and therefore does not contribute. In the asymptotically flat limit, it corresponds to the imaginary wormhole that saturates the imaginary distance bound (IDB). The same conclusion follows independently on the scalar side of the duality, and is confirmed by the exact amplitude. Replacing the Dirichlet condition with a one-parameter Neumann condition, the cross-throat saddle that leads to the imaginary wormholes remains irrelevant for the physical lift of the contour. This purely imaginary parameter characterizing the Neumann condition interpolates continuously between the half- and complete-wormhole geometries. Convergence of the sum over fixed-charge sectors imposes a corresponding interpolating bound not only on the imaginary part of the boundary axion but also on the parameter characterizing the Neumann boundary condition. An independent analysis based on the KSW criterion reproduces exactly the same bound, emphasizing its compatibility with the IDB.

\end{abstract}
\vspace{5mm}

%%%%%%%%%%%%%%%%%%%%%%%%%

\maketitle

\tableofcontents

%%%%%%%%%%%%%%%%%%%%%%%%%%%%%%%%%%%%%%%%%%%%%%%

%%%%%%%%%%%%%%%%%%%%%%%%%%%%%%%%%%%%%%%%%%%%%%%

\section{Introduction}
\label{sec:introduction}
\begin{figure}[hbtp]
    \centering
\includegraphics[width=0.6\linewidth]{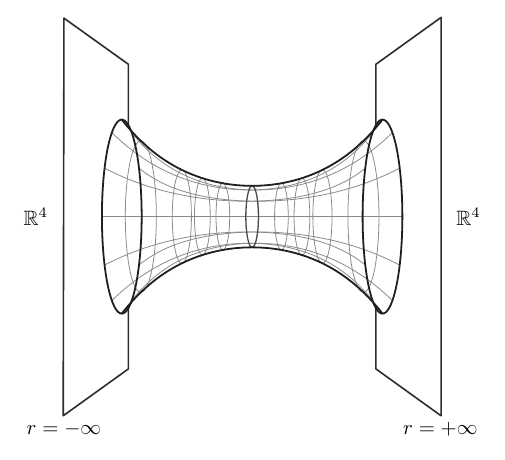}
    \caption{The Euclidean Giddings-Strominger(GS) wormhole connecting to asymptotically flat geometries ($\mathbb{R}^4$).}
    \label{fig:GS-wormhole}
\end{figure}

Spacetime wormholes play an important role in the study of the gravitational path integral. The prototypical example of such a wormhole is the Euclidean Giddings-Strominger (GS) wormhole \cite{Giddings:1987cg}, which connects two asymptotically flat spacetimes, see fig. (\ref{fig:GS-wormhole}). These are the solutions of Einstein's equation, with the stress-tensor given by the two-form gauge potential ($B$) with three-form flux ($H=dB$). Past studies have found that such wormholes are perturbatively stable and hence are valid candidate saddles in the path integral \cite{Loges:2022nuw, Jonas:2023ipa, Jonas:2023qle, Marolf:2025evo, Loveridge:2025dls, Hertog:2024nys, Hertog:2018kbz,Hebecker:2018ofv}. Recent studies \cite{Maldacena:2026jqd, DiUbaldo:2026rly} have found that such wormhole saddles provide important constraints on the properties of the low-energy effective field theory and imply an upper bound on the imaginary continuation of the boundary axion (IDB) consistent with the axionic weak gravity conjecture.

The analysis in \cite{Maldacena:2026jqd, DiUbaldo:2026rly} was performed assuming that the saddles contribute to the path integral. However, as the past studies show, the existence of a saddle doesn't always guarantee that it contributes to the path integral \cite{Held:2026huj,Held:2026bbo}. It depends on the boundary condition and on the integration cycle chosen in the complexified space of fields. Hence, the integration contour has to be the part of wormhole contribution. Furthermore, the saddle also has to satisfy the Kontsevich-Segal-Witten (KSW) criterion \cite{Kontsevich:2021dmb, Witten:2021nzp}. According to this criterion, the contributing saddles must be consistently coupled to all $p$-form gauge theories. For Euclidean wormholes, the criterion is always satisfied. Hence, one is prompted to check their contribution via the former.

The purpose of this work is twofold. The first is to investigate the allowability and Picard-Lefschetz relevance of Euclidean and complex wormhole saddles, focusing in particular on imaginary wormholes and the associated distance bound in the two dual formulations \cite{Witten:2026twr}. Because the Euclidean gravitational path integral suffers from the conformal-factor problem \cite{Gibbons:1976ue, Gibbons:1978ac}, we instead take the Lorentzian path integral as the starting point and analyze it using Picard-Lefschetz (PL) theory \cite{Witten:2010cx,Tanizaki:2014xba}. In the resulting complexified configuration space, Lorentzian, Euclidean, and genuinely complex saddles can be treated on an equal footing, with their contributions determined by the associated thimble decomposition. This framework has proved fruitful in a wide range of applications in quantum cosmology and black-hole physics \cite{Lehners:2023yrj, Feldbrugge:2017kzv, DiTucci:2019bui, Ailiga:2023wzl, Ailiga:2024wdx, Ailiga:2025osa,Ailiga:2025fny, Matsui:2021oio}.

Besides analyzing the compatibility of the two dual formulations and determining the relevance of their saddle points, the second goal of our work is to understand the compatibility between the imaginary distance bound with the KSW allowability criterion without reference to a particular integration contour. In \cite{Maldacena:2026jqd}, it was noticed that the IDB and KSW criteria coincide in some cases. Here, we aim to realize it explicitly within the minisuperspace framework. This provides a potentially distinct perspective from the first approach, which seeks to understand the distance bound in terms of the saddles that contribute to the path integral. To achieve it, we construct a one-parameter family of Euclidean wormholes interpolating between the half and complete GS wormholes. This is possible when we impose a particular boundary condition on the path integral.

In this work, we will be analyzing the following Lorentzian path integral
\begin{equation}
    \label{eq:path-integral}
    Z[{\rm Bd}_q(0),{\rm Bd_q(1)}]=\int_{\mathcal{C}_L}[\mathcal{D}g_{\mu\nu}\mathcal{D}\mathbf{m}] \exp\biggl[iS_{\rm grav}(g_{\mu\nu},\mathbf{m})/\hbar\biggr],
\end{equation}
where $\mathcal{C}_L$ defines the integration cycle, ${\rm Bd}_q(0)$ and ${\rm Bd}_q(1)$ are the boundary conditions on the metric $g_{\mu\nu}$. $\mathbf{m}$ denotes an additional matter field supporting the wormhole solution, which could be a two-form gauge field ($B$) or a dual scalar axion ($\theta$) \cite{Witten:2026twr}, and $S_{\rm grav}$ is the action for the theory consisting of gravity plus matter. We will analyse the above path integral in the context of Euclidean and complex wormhole saddles within the duality framework for the chosen boundary conditions. While the complete equivalence holds for any inhomogeneous background, we will restrict ourselves to the geometries respecting the following minisuperspace ansatz

\begin{equation}
\label{eq:metric}
ds^2 = -\frac{N_c^2}{q(t)}\,dt^2 + q(t)\,d\Omega_3^2,
\end{equation}
where $t\in[0,1]$, $d\Omega_3^2$ is the metric on the unit sphere, and the geometries have topology $\mathbb{R}\times S^3$. We impose the proper-time gauge on the lapse ($N_c$) \cite{Teitelboim:1983fk, Dasgupta:2001ue, Banihashemi:2024aal}. We also choose for both the axion and the three-form field strength to depend only on $t$ \cite{Loges:2022nuw}.

To study the allowability of the saddles, we start by considering the Dirichlet boundary condition, which fixes the metric. The saddle geometries in the Euclidean regime correspond to the various finite segments of the full GS wormholes. There are two physically distinct types of geometry- ``same side" and ``cross throat" corresponding to the wormholes whose boundaries lie on the same side and opposite side of the throat, respectively. Choosing the real lapse integration contour, the relevance of the saddles can be inferred from the PL technique. We analyse the relevance on both sides of the duality. The imaginary wormholes, as proposed in \cite{Maldacena:2026jqd, DiUbaldo:2026rly} can be obtained from the asymptotic analysis of the geometries. In the dual representation, the saddle corresponds to the imaginary distance bound (IDB).

It is worthwhile to ask how the saddles' relevance changes when we change the boundary condition. Concrete realizations of this idea arise in quantum cosmology \cite{Lehners:2023yrj, Ailiga:2023wzl, Ailiga:2024wdx, DiTucci:2019bui}, where changing the boundary conditions modifies the Picard-Lefschetz decomposition and, consequently, the set of relevant saddle points. We consider the Neumann condition on the metric and realize the same saddle geometries \cite{Lavrelashvili:2026jcw,Ailiga:2024wdx,Krishnan:2016mcj,DiTucci:2019bui}. By choosing the real integration contour, we study its relevance on both sides of the duality. An interesting feature of the Neumann condition is its invariance under the scaling symmetry of the flat-space wormhole. As a consequence, it yields a distinctive feature in the asymptotic regime. 

As explained in \cite{Witten:2026twr}, the complete quantum duality involves the sum over fixed charge-flux wormholes. While the sum is convergent for real values of axion, it can fail to converge if axion is allowed to take an arbitrary complex value. This in turn implies an upper bound on the imaginary part of the axion \cite{DiUbaldo:2026rly}. The scaling symmetry of Neumann together with the above convergence criterion of the sum allows us to obtain one parameter family of the bound on the imaginary part of the axion interpolating between the half and full imaginary wormhole. It is also important to see how the above bound compares with the Kontsevich-Segal-Witten criterion. Hence, we rederive the above bound from this criterion.

The remainder of this paper is organized as follows. In Sec.~\ref{sec:duality_and_condition}, we review the four-dimensional duality between a two-form gauge field and a scalar axion, paying particular attention to the metric boundary conditions and the lapse integration contour. In Sec.~\ref{sec:dirichlet-wormhole}, we impose Dirichlet boundary conditions on the metric and examine the allowability and relevance of the imaginary-wormhole saddles associated with the imaginary distance bound (IDB) in the two dual formulations. In Sec.~\ref{sec:Neumann-wormhole}, we extend the analysis to Neumann boundary conditions and study the relevance of the wormhole solutions on both sides of the duality. In Sec.~\ref{sec:ksw}, we apply the Kontsevich-Segal-Witten (KSW) allowability criterion and derive a distance bound consistent with the convergence condition for the flux sum. This shows the compatibility between KSW and the IDB bound. Finally, Sec.~\ref{sec:conclusion} summarizes our main results and discusses their implications.

%%%%%%%%%%%%%%%%%%%%%%%%%%%%%%%%%%%%%%%%%%%%%%%%%%%%%%%%%%%%%%%%%

\section{The duality and the boundary condition}
\label{sec:duality_and_condition}

In this work, we consider the duality between the massless scalar axion ($\theta$) and the two-form field ($B$) with the field strength $H=dB$, in four spacetime dimensions \cite{Giddings:1987cg, Witten:2026twr}. The actions for the two sides of duality are related through
\begin{equation}
    \label{eq:duality}
    \frac{1}{2 f_\theta^2}\int d^4 x\sqrt{-g}\,\,\partial_\mu\theta\partial^\mu\theta\leftrightarrow \frac{1}{2.3!f_h^2}\int d^4x\sqrt{-g}\,\,H^{\mu\nu\rho}H_{\mu\nu\rho},
\end{equation}
where $f_\theta$ and $f_h$ are the couplings, $g_{\mu\nu}$ is the background metric and $\sqrt{-g}$ is the determinant of the metric. The duality requires the Dirichlet condition to be imposed on the axion and the Neumann condition to be imposed on the 3-form field. It also requires appropriate constraints on the couplings and quantization condition. The duality holds for any background metric $g_{\mu\nu}$, and thus it is argued to hold when we couple both sides of eq. (\ref{eq:duality}) with the gravity action \cite{Witten:2026twr}. Under the duality, the stress-tensor and the equations of motion for these two theories become identical (see appendix \ref{sec:stres_tensor_equality} for more details). When gravity becomes dynamical, one introduces two important key ingredients into the duality- the boundary condition on the metric and the contour of integration. In this section, we briefly examine the duality, focusing on two aspects within the Lorentzian path-integral formalism. A similar setup has also been considered in \cite{Loges:2022nuw}. While the complete equivalence holds for any inhomogeneous background, we will restrict ourselves to the geometries respecting the metric ansatz in eq. (\ref{eq:metric}) %following minisuperspace ansatz

Let us first consider the three-form side of the duality. After coupling to E-H gravity, the total action is given by
\begin{equation}
\label{eq:3-form_action}
    S_{3-{\rm form}}[g_{\mu\nu},H]=\frac{1}{16\pi G}\int d^4x\sqrt{-g} \,R-\frac{1}{2.3! f_h^2}\int d^4x\sqrt{-g}\,\,H^{\mu\nu\rho}H_{\mu\nu\rho}+S_{\rm bd},
\end{equation}
where $S_{\rm bd}$ is the boundary term required to achieve proper variation depending on the boundary condition. By choosing an appropriate boundary condition, one can obtain a background that could be Lorentzian, Euclidean, or even complex. Much of our discussion will focus on Euclidean wormholes. Of particular importance is a special class of backgrounds, commonly referred to as ``imaginary wormholes", \cite{Maldacena:2026jqd, DiUbaldo:2026rly}. We choose the three-form field as $H_{\mu\nu\rho}=h(t)\ep_{\mu\nu\rho}$, $\ep_{\mu\nu\rho}$ being the Levi-Civita tensor on unit $S^3$. Imposing the Neumann condition on the 3-form, $h(t)$ becomes constant (and integer quantized) and doesn't require any surface terms in the variation. For the background, we denote the generic boundary conditions to be ${\rm Bd_q(0)}$ and ${\rm Bd_q(1)}$ at $t=0$ and $t=1$ respectively. Hence, $S_{\rm bd}$ is the surface terms arising from the metric part only. The Lorentzian path integral for the geometries in eq. (\ref{eq:metric}) and the gravity action in eq. (\ref{eq:3-form_action}) is given by ($8\pi G=1$)
\begin{equation}
    \label{eq:path_integral_3form}
    \begin{split}
    &Z_n[{\rm Bd_q(0)},{\rm Bd_q(1)}]=\int_0^\infty dN_c\int \mathcal{D}q \exp[iS_{3-\rm form}[q,N_c,h]]\\
    \text{where,}\quad& S_{3-\rm form}[q,N_c,h]/2\pi^2=\int_0^1 dt \biggl[3N_c-\frac{3\dot{q}^2}{4N_c}-\frac{h^2N_c}{2f_h^2q^2}\biggr]+\biggl[\frac{3q\dot{q}}{2N_c}\biggr]_0^1+S_{\rm bd}[q,\dot{q}].
    \end{split}
\end{equation}
Let us turn to the scalar axion ($\theta$) side of the duality. The action for EH gravity coupled to a massless axion is given by
\begin{equation}
    \label{eq:action_axion}
    S_{\rm axion}[g_{\mu\nu},\theta]=\frac{1}{16\pi G}\int d^4x\sqrt{-g} \,R-\frac{1}{2f_\theta^2}\int d^4x\sqrt{-g}\,\,\partial^\mu\theta\partial_\mu\theta+S_{\rm bd}.
\end{equation}
For the Dirichlet condition on the axion, we require no additional surface terms. Owing to the shift symmetry of the massless axion $\theta\rightarrow\theta+c$, $c$ being a constant, we choose the boundary condition on the axion to be the following
\begin{equation}
    \label{eq:bc_axion}
    \theta(0)=0,\quad\theta(1)=\Delta\theta.
\end{equation}
We choose the boundary value of the axion on the initial hypersurface to vanish. $\Delta\theta$ describes the difference between the boundary values of the axion, which plays a crucial role in our discussion in the context of imaginary wormholes. The Lorentzian path integral for the metric in eq. (\ref{eq:metric}) and the gravity action in eq. (\ref{eq:action_axion}) is given by ($8\pi G=1$)
\begin{equation}
    \label{eq:path_integral_axion}
    \begin{split}
    &Z_{\rm axion}[{\rm Bd_q(0)},{\rm Bd_q(1)};\Delta\theta]=\int_0^\infty dN_c\int \mathcal{D}q \int\mathcal{D}\theta\exp[iS_{\rm axion}[q,N_c,\theta]]\\
   \text{where,}\quad & S_{\rm axion}[q,\theta,N_c]/2\pi^2=\int_0^1 dt \biggl[3N_c-\frac{3\dot{q}^2}{4N_c}+\frac{q^2\dot{\theta}^2}{2f_\theta^2 N_c}\biggr]+\biggl[\frac{3q\dot{q}}{2N_c}\biggr]_0^1+S_{\rm bd}[q,\dot{q}].
    \end{split}
\end{equation}
Since the path-integral over the $\theta$ is quadratic, it can be done exactly, yielding
\begin{equation}
    \label{eq:axion-path-integral-exact}
    \begin{split}
&\int_{\theta(0)=0}^{\theta(1)=\Delta\theta} \mathcal{D}\theta \exp\biggl[2\pi^2i\int_0^1dt \frac{q^2\dot{\theta}^2}{2f_\theta^2N_c}\biggr]=\sqrt{\frac{\pi}{if_{\theta}^2T[q]}}\exp\biggl[\frac{i\pi^2(\Delta\theta)^2}{T[q]f_\theta^2}\biggr],\quad T[q]=N_c\int_0^1 \frac{ds}{q(s)^2}.
    \end{split}
\end{equation}
The quantity multiplied by the exponential is the one-loop factor that arises from the homogeneous modes of the scalar field. Within the minisuperspace setup, these are the only modes that contribute to the path integral. Note, we keep the boundary condition on the metric generic. Since the axion and the three-form field are Fourier conjugate, we take the fourier transformation of $Z_{\rm axion}[{\rm Bd_q(0)},{\rm Bd_q(1)};\Delta\theta]$ with $e^{-i\Delta\theta n},\,n=2\pi^2 h,\,\, n\in\mathbb{Z}$\footnote{The presence of the factor $e^{in\Delta\theta}$ has been argued physically in \cite{Witten:2026twr} as a choice of vacuum in non-compact geometries.}. Performing the following Fourier integral over $\Delta\theta$, we get
\begin{equation}
    \label{eq:identity}
    \int_{-\infty}^{+\infty}d(\Delta\theta) \exp\biggl[i\Delta\theta\, n+\frac{i\pi^2(\Delta\theta)^2}{T[q]f_\theta^2}\biggr]=\sqrt{\frac{if_\theta ^2T[q]}{\pi}}\exp\biggl[-\frac{i n^2f_\theta^2 T[q]}{4\pi^2}\biggr].
\end{equation}
Note that the one-loop contribution arising from the homogeneous modes of the scalar precisely cancels the identical factor that arises in the Fourier transformation. In case when the couplings are related $f_\theta=1/f_h$ \cite{Witten:2026twr}, it yields the following identity
\begin{equation}
\label{eq:fourier_transform}
    \begin{split}
        Z_n[{\rm Bd_q(0)},{\rm Bd_q(1)}]=\int_{-\infty}^{+\infty}d(\Delta\theta) e^{i\Delta\theta\, n}Z_{\rm axion}[{\rm Bd_q(0)},{\rm Bd_q(1)};\Delta\theta].
    \end{split}
\end{equation}
The above equation holds true for any generic boundary condition on the metric ${\rm Bd}_q$. background ($q$). This will become more evident once we do the saddle analysis on both sides of eq. (\ref{eq:fourier_transform}). In our paper, we study it for two types of boundary conditions- Dirichlet and Neumann, focusing on the Euclidean wormholes and complex saddles geometries.  

Let us consider the boundary conditions which describe the wormhole background as shown in fig \ref{fig:GS-wormhole}. The duality requires $n\in\mathbb{Z}$ to be quantized and correspondingly the conjugate axion needs to be compact with $\theta\sim\theta+2\pi$. The eq. (\ref{eq:fourier_transform}) is expressed in the covering space of the axion, which is taken all the way from $-\infty$ to $+\infty$. The complete duality involves summing over all the discrete $n$ \cite{Witten:2026twr}. This implies summing over fixed-flux wormholes 
\begin{equation}
    \label{eq:n_sum_identity}
\Psi[\Delta\theta]=\sum_{n\in\mathbb{Z}}e^{in\Delta\theta}Z_n.
\end{equation}
The above quantity is often called the fixed-angle representation, in which we fix the angle $\Delta\theta$, the boundary value of the axion. $Z_n$ is an even function of $n$. The sum plays a central role in our discussions, as the requirement of its convergence yields the imaginary distance bound (IDB).

%%%%%%%%%%%%%%%%%%%%%%%%%%%%%%%%%%%%%%%%%%%%%%%%%%%%

\section{Wormholes with Dirichlet condition}
\label{sec:dirichlet-wormhole}
We start by considering the Dirichlet boundary conditions on the metric at $t=0$ and $t=1$: $q(0)=q_0, q(1)=q_1$. We allow $q_0$ and $q_1$ to be large, but below some large cutoff ($R$). A similar set-up has been considered in \cite{Loges:2022nuw}. Our analysis, however, extends beyond the scope of these earlier works, as will become clear as we go along. We start by computing the path integral exactly for the three-form side of the duality, followed by the saddles and Picard-Lefschetz analysis. Then we discuss about the `` imaginary wormholes" by analyzing asymptotic behaviour of the saddle geometries. Finally, we perform the analysis in the scalar side of the duality. 

%%%%%%%%%%%%%%%%%%%%%%%%%%%%%%%%%%%%%%%%%%%%%%%

\subsection{The exact fixed-flux path integral}
\label{sec:exact_flux_integral_dirchlet}
In this section, we compute the path integral exactly for the fixed three-form side of the duality as given in eq. (\ref{eq:path_integral_3form}). Taking the appropriate GHY surface term, the path integral takes the following form ($f_h=1,8\pi G=1,n=2\pi^2h$)
\begin{equation}
    \label{eq:fixed_flux_path_inte}
    \begin{split}
    &Z_n[q_0,q_1]=\int_0^\infty dN_c \int\mathcal{D}q \exp\biggl[iS_{\rm 3-form}(q,N_c,n)\biggr]\\
    \text{where}\quad &S_{\rm 3-form}(q,N_c,n)=2\pi^2\int_0^1 dt\biggl[3N_c-\frac{3\dot{q}^2}{4N_c}-\frac{n^2N_c}{8\pi^4q^2}\biggr],
    \end{split}
\end{equation}
where we have added $S_{\rm bd}=-3q\dot{q}/2N_c$ at the two boundaries to achieve proper variation.
The integral above over the background ($q$) can be evaluated exactly. However, as the integrand is singular at $q=0$, it needs additional care. One requires additional UV information near $q=0$, or a suitable self-adjoint extension. However, those prescriptions will be insensitive for the purposes of this paper. We follow the simplest Feynman prescription to compute the path integral, which is analytically continued radial time sliced kernel and is not self-adjoint, see appendix \ref{sec:exact-path-integral} for details. Performing the path integral over $q(t)$ along with the $N_c$ integral exactly \cite{Grosche:1998yu}, we get (see the appendix \ref{sec:exact-path-integral} for derivation) 
\begin{equation}
\label{eq:fixed_flux_path_integral_exact}
    Z_n[q_0,q_1]=\frac{\sqrt{q_0q_1}}{\hbar}I_\nu\biggl(\frac{6\pi^2}{\hbar}q_<\biggr)K_\nu\biggl(\frac{6\pi^2}{\hbar}q_>\biggr),\quad \nu=\sqrt{\frac{1}{4}-\frac{3n^2}{2\hbar^2}},
\end{equation}
where, $q_<=\text{min}(q_0,q_1),q_>=\text{max}(q_0,q_1)$, $I_\nu$ and $K_\nu$ are the modified Bessel functions. The path integral is restricted to the positive half plane $q(t)>0$. The $1/4$ inside the $\nu$ is the Langer shift arising from the quantum correction. Having obtained the exact expression, we take the semiclassical limit, which helps us to identify the saddle contributions essential for the Picard-Lefschetz analysis in the next section. Since $Z_n=Z_{-n}$, it is useful to define $\tilde{n}=|n|/2\pi^2\sqrt{6}$. As our focus is primarily on the Euclidean wormholes, we consider the regime $\tilde{n}<q_0<q_1$. Utilizing the following uniform expansions of Bessel's function of large order and argument \cite{DLMF}, we get
\begin{equation}
    \label{eq:asymptotic_bessel}
    \begin{split}
        &K_{i\mu}(\mu z)\sim \sqrt{\frac{\pi}{2\mu}}\frac{e^{-\pi\mu/2-\mu g_2(z)}}{(z^2-1)^{1/4}}(1+\mathcal{O}(1/\mu))\\
        &I_{i\mu}(\mu z)\sim \frac{e^{\pi\mu/2}}{\sqrt{2\pi\mu}(z^2-1)^{1/4}}\biggl[e^{\mu g_2(z)}(1+\mathcal{O}(1/\mu))-\underbrace{\frac{i}{2}e^{-\mu g_2(z)}(1+\mathcal{O}(1/\mu))}_{\text{exponentially improved}}\biggr],
    \end{split}
\end{equation}
where $g_2(z)=\sqrt{z^2-1}-\sec^{-1}z$ and $|z|>1$. The second term in the asymptotic expansion of $I_{i\mu}(\mu z)$ is exponentially improved and cannot be captured in the ordinary Poincaré-type expansion. This term is required to satisfy the exact connection formula $I_{i\mu}(\mu z)-I_{-i\mu}(\mu z)=2K_{i\mu}(\mu z)\sinh(\pi\mu)/i\pi$. Utilizing the eq. (\ref{eq:asymptotic_bessel}), we observe that the $Z_n$ can receive contributions only from the following exponents
\begin{equation}
    \label{eq:exponent_exact_expression}
    Z_n\sim e^{6\pi^2\tilde{n}(- g_2(q_>/\tilde{n})+ g_2(q_</\tilde{n}))/\hbar}\quad \text{and/or}\quad e^{-6\pi^2\tilde{n}(g_2(q_>/\tilde{n})+ g_2(q_</\tilde{n}))/\hbar}.
\end{equation}
Note that one of the two terms in the exponent is always negative. In the regime when either $\tilde{n}>q_0$ or $q_1<\tilde{n}$ one obtains oscillatory phase, see appendix \ref{sec:exact-path-integral}. As explained there, the self-adjoint extension that we ignored doesn't introduce any new exponential structure. In the next section, we will identify the above contributions to the different kinds of saddle geometries and perform the Picard-Lefschetz analysis. 

%%%%%%%%%%%%%%%%%%%%%%%%%%%%%%%%%%%%%%%
 
\subsection{Saddles and Picard Lefschetz relevance}
\label{sec:saddle_and_PL}
\begin{figure}%[hbtp]
    \centering
   \subfigure[\,$\frac{1}{\tilde{n}^2}<\frac{1}{q_1^2}+\frac{1}{q_0^2}$\,]{ \includegraphics[width=0.48\linewidth]{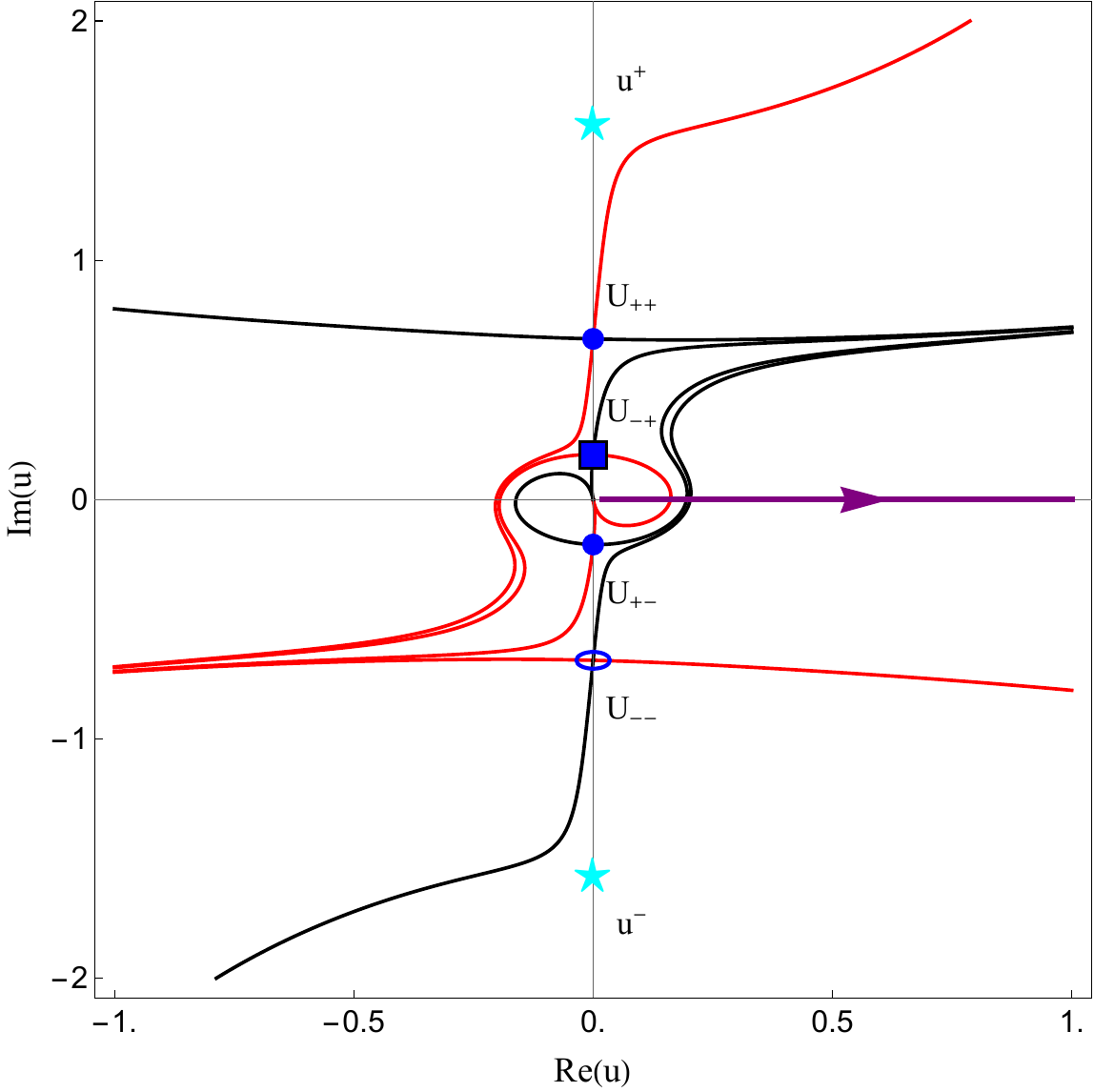}}
   \subfigure[\,$\frac{1}{\tilde{n}^2}>\frac{1}{q_1^2}+\frac{1}{q_0^2}$\,]{ \includegraphics[width=0.48\linewidth]{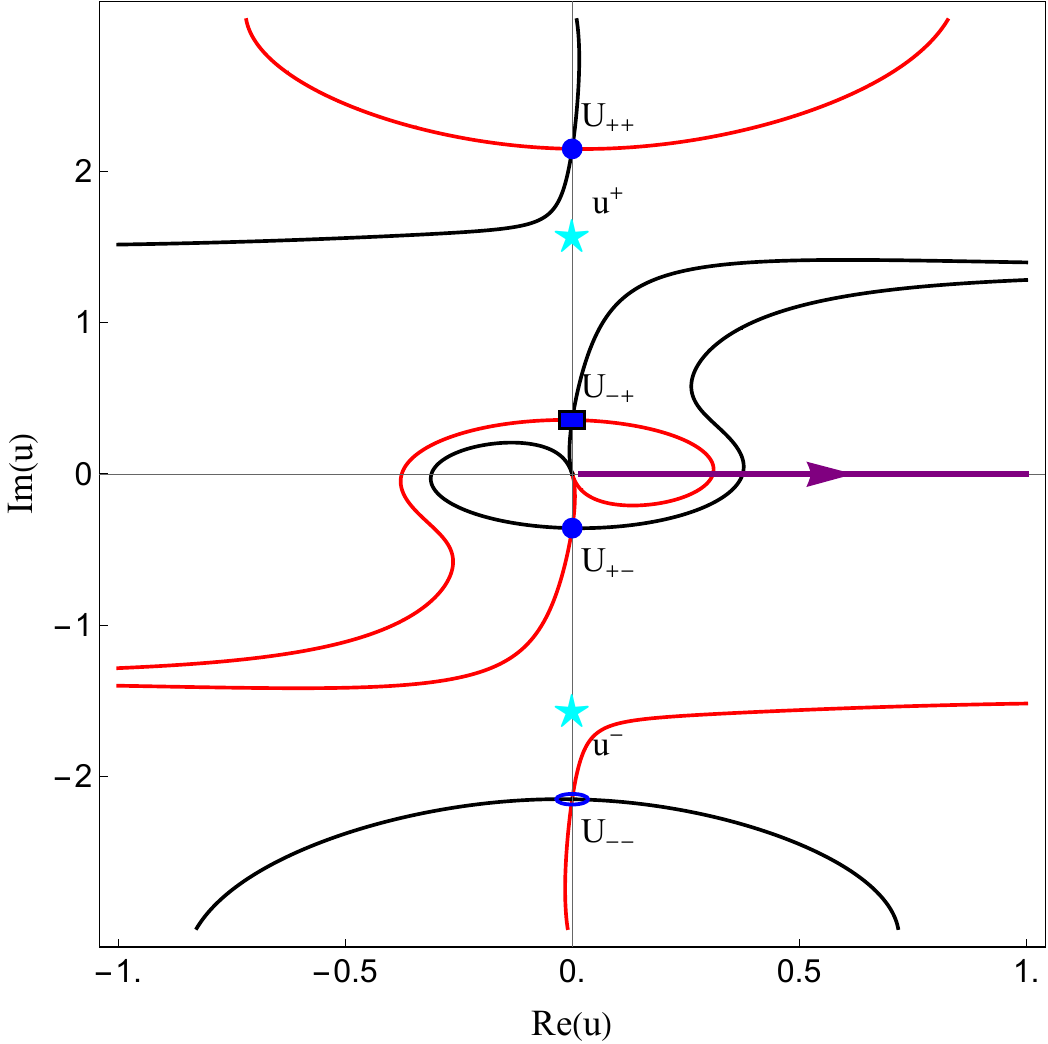}}
\caption{PL plot in the complex 
  $u$-plane. We choose $q_0 =1.03,$ $q_1=1.1,$ $\tilde{n}=1,$ $n=48$  for fig (a) and $q_0 =0.4,$ $q_1=0.8,$ $\tilde{n}=0.25,$ $n=12$ for fig (b). Steepest descent (black) and ascent (red) lines. Blue markers are the ramification points. The purple arrow denotes the integration cycle $\mathcal{C}_u=(0^{+},\infty)$. $U_{--}$ and $U_{++}$ are the cross-throat saddles, which are irrelevant; hence, the intersection number is $n_\sigma=0$. The relevant saddle $U_{-+}$ is depicted by a blue square. To break the stokes degeneracy, we rotate $G=|G|e^{i\ep},\ep=\pi/10$.}
\label{fig:PL_plot_euclidean_dirichlet}
\end{figure}

To determine the saddle points and their associated geometries, we solve the classical equation of motion derived from the action in eq. (\ref{eq:fixed_flux_path_inte}). It is
\begin{equation}
    \label{eq:on-shell}
    \ddot{q}+\frac{4\tilde{n}^{2}N_{c}^{2}}{q^{3}}=0.
\end{equation}
The solution to the above equation, subject to the boundary conditions $q(0)=q_0$ and $q(1)=q_1$, is
\begin{equation}
    \label{eq:q(t)_dirichlet}
    q_\pm(t)=\sqrt{q_0^2(1-t)^2+q_1^2t^2+2 q_0q_1 \Gamma_\pm t(1-t)},\qquad \Gamma_\pm=\pm\sqrt{1+\frac{4\tilde{n}^2N_c^2}{q_0^2q_1^2}}.
\end{equation}
The on-shell solution has two sheets. The negative branch (``$-$") is unphysical because, on the real $N_c$ contour, $q_-(t)$ passes through zero, rendering the path integral ill-defined. We always assume $q(t)>0$ consistent with the exact path integral definition. Evaluating the on-shell action from eq. (\ref{eq:fixed_flux_path_inte}), the path integral reduces to an ordinary integral over $N_c$
\begin{equation}
    \label{eq:lapse-action}
    \begin{split}
& Z_n[q_0,q_1]=\int_0^\infty dN_c \exp(iS[N_c])\\
\text{where},\quad & S[N_c]=2\pi^2\biggl[3N_c-\frac{3(q_0^2+q_1^2)}{4N_c}+\frac{3q_0q_1}{2N_c}f\biggl(\frac{2\tilde{n}N_c}{q_0q_1}\biggr)\biggr],
\end{split}
\end{equation}
where $f(z)=\sqrt{z-i}\sqrt{z+i}-z\ln(z+\sqrt{z-i}\sqrt{z+i})$. The action has branch points at $N_c^*=\pm i q_0q_1/2\tilde{n}$. The saddles of the action are
\begin{equation}
    \label{sec:saddle_dirichlet}
    N_{\ep_1,\ep_2}=\frac{i}{2}\biggl(\ep_1\sqrt{q_0^2-\tilde{n}^2}+\ep_2\sqrt{q_1^2-\tilde{n}^2}\biggr),\quad \ep_1,\ep_2=\pm.
\end{equation}
To determine the geometries at these saddle points, we substitute the corresponding solutions into the metric ansatz in Eq. (\ref{eq:metric}). Upon writing $q=a^2$, a straightforward calculation yields
\begin{equation}
    \label{eq:Giddings-stominger-wormholes}
    ds^2=\frac{da^2}{1-(\frac{a_{\rm th}}{a})^4}+a^2d\Omega_3^2, \qquad a_{\rm th}^2=\tilde{n}.
\end{equation}
The above geometry is the familiar metric for Giddings-Strominger (GS) wormholes connecting two asymptotically flat boundaries \cite{Giddings:1987cg}, where the throat of the wormhole ($a_{\rm th}$) is supported by $\tilde{n}$, see fig. \ref{fig:GS-wormhole}. Although $\ep_1,\ep_2$ drops from the metric element, they represent different segments of the full geometry. In the regimes where $\tilde{n}<q_0<q_1$, it describes a Euclidean geometry (imaginary saddles). For other parameter values, the geometry becomes complex. To analyze the saddles, we call the saddles as ``same-side" and "cross-throat ", as follows
\begin{equation}
    \label{eq:sad-cross-same}
    N_{\rm same-side}: N_{+-}\,\, \&\,\, N_{-+},\quad N_{\rm cross-throat}: N_{++}\,\, \&\,\, N_{--}.
\end{equation}
\begin{figure}[hbtp]
    \centering
    \includegraphics[width=1\linewidth]{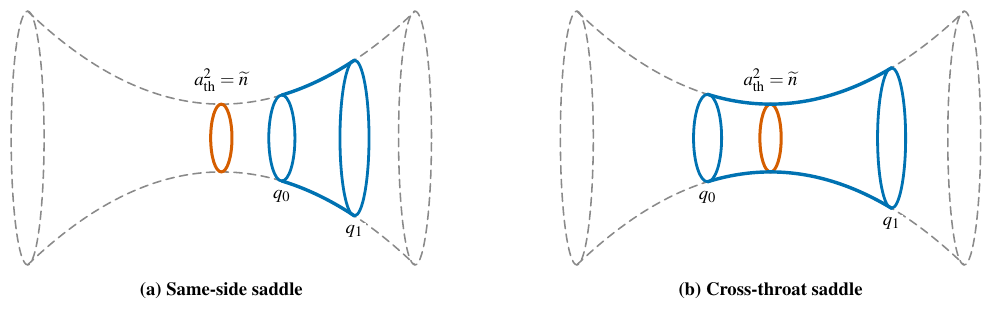}
    \caption{Two types of the Euclidean Wormhole geometry for the range of parameter $\tilde{n}<q_0<q_1$, where $a_{\rm th}$ is the size of the throat of the wormhole. $N_{+-}$ and $N_{-+}$ represent the same side wormhole geometry while $N_{--}$ and $N_{++}$ represent the cross-throat wormhole saddle. The grey dashed background is the full wormhole profile, while the blue portion describes the saddle segment. When we take $q_0$ and $q_1$ large, both hypersurfaces in the same-side configuration approach the same (right) asymptotic region. By contrast, in the cross-throat configuration, the two hypersurfaces approach opposite asymptotic regions, thereby recovering the complete two-sided wormhole geometry.}
    \label{fig:saddle-geometry}
\end{figure}
For the saddles denoted by $N_{\rm cross-throat}$, the background profile can generate a minima/throat at finite time ($t^*$) while the ``same-side" saddles are monotonic, see fig. \ref{fig:saddle-geometry}. For $\ep_1\ep_2=-1$, the two boundaries lie on the same side of the throat, corresponding to the same-side geometry. By contrast, $\ep_1\ep_2=+1$ places the boundaries on opposite sides of the throat, yielding the cross-throat geometry. It can be seen explicitly by examining $q(t)$ at these saddles
\begin{equation}
    \label{eq:q(t)_at_saddle}
    \begin{split}
    \ep_1\ep_2=-1,\qquad &q(t)\biggl|_{N_{\ep_1\ep_2}}=\{\tilde{n}^2+[A_0+(A_1-A_0)t]^2\}^{1/2}\\
    \ep_1\ep_2=1,\qquad & q(t)\biggl|_{N_{\ep_1\ep_2}}=\{\tilde{n}^2+[A_0-(A_1+A_0)t]^2\}^{1/2},
    \end{split}
\end{equation}
where we defined $A_0=\sqrt{q_0^2-\tilde{n}^2},A_1=\sqrt{q_1^2-\tilde{n}^2}$. For the cross-throat saddles, $q(t)$ hits the throat at $t^*=A_0/(A_0+A_1)$. It is also important to note that for the parameters satisfying
\begin{equation}
    \label{eq:branch-condition}
    \frac{1}{\tilde{n}^2}<\frac{1}{q_1^2}+\frac{1}{q_0^2},
\end{equation}
the cross-throat saddles ($\ep_1\ep_2=1$) remains in the $+$ branch. When the inequality is not satisfied, the saddles move to the other branch. The ``same side" saddles ($\ep_1\ep_2=-1$) can never be pushed behind the branch cut and remain always in the $+$ branch. This directly follows from observing that $|N_{+-}|=|N_{-+}|<|N_c^*|,\forall \,\, q_0,q_1.$ As a consequence of eq. (\ref{eq:branch-condition}), we note that it is always violated for sufficiently large values of $q_0$ and $q_1$. Picard Lefschetz analysis has been done in the $N_c$-plane in different regimes of parameters \cite{Loges:2022nuw}. Here, our analysis will focus primarily on the Euclidean saddles, which will play an important role in our discussion of imaginary wormholes. As shown in \cite{Witten:2010cx}, working in a covering plane ($u$), one can avoid the cuts:
\begin{equation}
\label{eq:covering_map_dirichlet}
    N\rightarrow u,\quad N=q_0 q_1\sinh(u)/2\tilde{n}.
\end{equation}
Utilising the map, one can identify the images of saddles and contour in the $u$-plane. We will work in the $u$-variable to figure out the relevance of the saddles. The off-shell $N_c$ action as given in eq. (\ref{eq:lapse-action}), in the $u$-variable becomes  

\begin{equation}
 \label{eq:action at u plane} 
S[u]=
2\pi^2\biggl[3\tilde{n}\,(\coth u-u)
-\frac{3\tilde{n}\left(q_0^2+q_1^2\right)}{2q_0q_1 \sinh u}
+\frac{3q_0q_1}{2\tilde{n}}\,\sinh u\biggr].
\end{equation}
The image of saddles in the $ u$-plane corresponding to the $N_c$-plane saddles mentioned in eq. (\ref{sec:saddle_dirichlet}) are

\begin{equation}
\label{eq:saddle_u_plane}
U_{\ep_1,\ep_2}
=
2\pi i\,k
+ \biggl[\ep_1\cosh^{-1}\biggl(\frac{\tilde{n}}{q_0}\biggr)+\ep_2\cosh^{-1}\biggl(\frac{\tilde{n}}{q_1}\biggr)\biggr],\quad \ep_1,\ep_2=\pm 1, k\in\mathbb{Z}
\end{equation}
The values $u_\pm=\pm\frac{i\pi}{2}+2\pi i\,k,\,\,k\in\mathbb{Z}$, which appear as saddles in the $u$-plane, are the ramification points, where the covering map degenerates. The integral in eq. (\ref{eq:lapse-action}) becomes
\begin{equation}
     \label{eq:u-plaane integral}
    Z_n[q_0,q_1]= \int_{\mathcal{C}_u} du \,\mathcal{A}(u)\,\exp(iS[u]),\qquad \mathcal{C}_u=(0^+,\infty).
 \end{equation}
The points where $\mathcal{A}(u)=q_0q_1\cosh(u)/2\tilde{n}$ vanish are the ramification points; the measure of the integration becomes zero. Since the saddles are all Euclidean and hence lying on the Stokes ray, we break the degeneracy by rotating Newton's constant ($G_N$) by a small complex phase $G=|G|e^{i\ep},\ep>0$ is chosen to ensure the convergence at large $u$ \cite{Ailiga:2025fny, Ivo:2025yek}. The Picard Lefschetz plots are shown in the fig. \ref{fig:PL_plot_euclidean_dirichlet} reveals that the saddle $N_{--}$ is always irrelevant in the computation of the path integral for the choice of contour $N_c\in (0^+,\infty)$ and its intersection number ($n_\sigma$) is zero. The steepest ascent from the $N_{--}$ saddle can never approach the contour of integration.   
The saddles corresponding to the ramification points $u_\pm$ are irrelevant in the path integral. 

%%%%%%%%%%%%%%%%%%%%%%%%%%%%%%%%%%%%%%%%%%%%%%%%
%%%%%%%%%%%%%%%%%%%%%%%%%%%%%%%%%%%%%%%%%%%%%%%%

\subsection{Asymptotic limit and the Imaginary wormhole}
\label{sec:asymptotic_limit_imaginary_wormhole}

\begin{figure}
    \centering
    \includegraphics[width=0.65\linewidth]{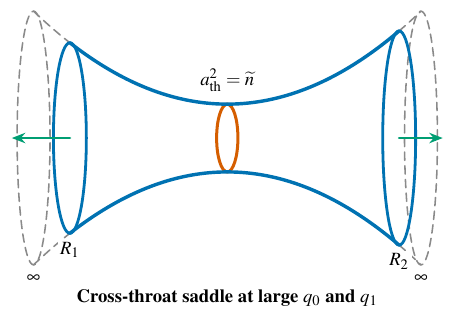}
    \caption{The picture shows that as we take large $q_0$ and $q_1$, the cross-throat saddle envelops the complete wormhole profile. Such rescaling of $q_0$ and $q_1$ doesn't alter the PL relevance in the regime $\tilde{n}<q_0<q_1$, as explained in the text. Performing the background subtraction, the action at the saddle agrees with the imaginary wormhole saddle satisfying IDB. }
    \label{fig:asympto_cross_throat}
\end{figure}
Having identified the relevant and irrelevant saddles of the path integral, we now investigate their geometric interpretation in the large-boundary limit. This is an extension to the discussion carried in \cite{Loges:2022nuw}. In particular, we seek to determine which saddles in the lapse plane correspond to the imaginary wormhole configurations. To identify these saddles, we examine the large-$q_0,q_1$ asymptotics of their respective on-shell actions. This limit describes the asymptotically flat regime. We restrict our analysis to the parameter range $\tilde{n}<q_0<q_1$, for which Euclidean wormhole saddles exist. As $q_0,q_1$ are taken to infinity, the condition is automatically satisfied for every fixed value of the three-form flux. Consequently, the upper bound on $\tilde{n}$ is removed in the asymptotically flat limit.
In this regime, the action at the four saddles mentioned in eq. (\ref{sec:saddle_dirichlet}) is given by
\begin{equation}
    \label{eq:saddle_at_action}
iS[N_{\ep_1,\ep_2}]=-6\pi^2\tilde{n}\biggl[\ep_1 g_2\biggl(\frac{q_0}{\tilde{n}}\biggr)+\ep_2 g_2\biggl(\frac{q_1}{\tilde{n}}\biggr)\biggr],
\end{equation}
where $g_2(z)=\sqrt{z^2-1}-\sec^{-1}z$. Since $g_2(z)$ is positive definite for $z>1$, the saddle corresponding to $\ep_1=\ep_2=-1$ yields the dominant contribution to the path integral.
We observe that if we take $q_0$ and $q_1$ asymptotically large, the action can diverge. Such a behaviour is expected in an asymptotically flat spacetime. Utilizing the asymptotic expansion $g_2(z)=z-\pi/2+\mathcal{O}(1/z)$, we get
\begin{equation}
    \label{eq:asymptotic behaviour}
    iS[N_{\ep_1,\ep_2}]\biggr|_{\{q_0,q_1\}\rightarrow \{R_1,R_2\}}=-6\pi^2\biggl[\ep_1 R_1+\ep_2 R_2-\tilde{n}(\ep_1+\ep_2)\frac{\pi}{2}+\mathcal{O}\biggl(\frac{1}{R_1},\frac{1}{R_2}\biggr)\biggr],
\end{equation}
where $R_1<R_2$ are the large cut-off at the asymptotic boundaries for the left and the right of the wormhole geometry. For $\ep_1\ep_2=-1$, both sides of the geometry extend toward the right, whereas for $\ep_1\ep_2=1$, two sides extend in opposite directions, with one approaching the left and the other the right asymptotics, as illustrated in (\ref{fig:saddle-geometry}). The divergent part in all cases is proportional to the cut-offs. Such divergences in asymptotically flat space can be renormalized by background subtraction \cite{Gibbons:1976ue}\footnote{The divergence at each hypersurfaces has the structure of $\int_{S^3(R_i)} d^3x \sqrt{\gamma}\, K_0$, where $i=1, 2$, $\gamma$ is the three metric and $K_0$ is the extrinsic curvature of $S^3$ embedded in the euclidean $\mathbb{R}^4$. It is commonly known as background subtraction \cite{Gibbons:1976ue}. Such a subtraction doesn't affect the thimbles in the complex $N_c$-plane as the term is independent of $N_c$. Indeed, as shown in appendix \ref{sec:symmetry}, such rescaling of $q_0$ and $q_1$ doesn't change the intersection number when we are in the Euclidean regime.}. Subtracting the divergences, we are left with the following renormalized part
\begin{equation}
    \label{eq:renorm_action}
    iS[N_{\ep_1,\ep_2}]_{\rm ren}=6\pi^2\tilde{n}(\ep_1+\ep_2)\frac{\pi}{2}.
\end{equation}
It is crucial to note that the action scales linearly with $\tilde{n}$. For $\ep_1\ep_2=-1$, the renormalized action vanishes as both the hypersurfaces coincide. However, for the choice, $\ep_1=\ep_2=-1$, we get (using $\tilde{n}=|n|/2\pi^2\sqrt{6}$)
\begin{equation}
    \label{eq:mmm_saddle}
    iS[N_{-,-}]_{\rm ren}=-\sqrt{\frac{3}{2}}\pi|n|,\qquad\Rightarrow\,\, Z_n\sim n_\sigma\exp(-\sqrt{3}\pi|n|/\sqrt{2}).
\end{equation}
Note that the renormalized action is negative and also is linear in $|n|$. The coefficient multiplied by $|n|$ is precisely twice the imaginary distance bound (IDB) as discussed in \cite{Maldacena:2026jqd, DiUbaldo:2026rly}. Hence, the saddle $N_{--}$ corresponds to the imaginary wormhole in the asymptotically flat limit. For the other choice of sign $\ep_1=\ep_2=1$, one gets a positive sign in the action. The contribution from $N_{--}$ saddle in the path will be the exponential of the renormalized action multiplied by the intersection number ($n_\sigma$), dictating relevance. Hence, $Z_n\sim n_\sigma \exp(-\sqrt{3}\pi|n|/\sqrt{2})$. However, as we have shown in the previous section, the saddle $N_{--}$ corresponds to an irrelevant saddle ($n_\sigma=0$) and therefore doesn't contribute to the path integral for the real lapse contour. Interestingly, the contribution from this saddle is also absent in the exact expression, as discussed in sec. \ref{sec:exact_flux_integral_dirchlet}. This is one of the claims of our paper.

To ensure completeness, let us consider the sum over a fixed charged wormhole as defined in eq. (\ref{eq:n_sum_identity}) for the saddle $N_{--}$. Removing the $n_\sigma$, we obtain
\begin{equation}
    \label{eq:flux_sum}
\Psi[\Delta\theta]=\sum_{n\in\mathbb{Z}}e^{in\Delta\theta}Z_n=\sum_{n\in\mathbb{Z}}\exp\biggl(-|n|\sqrt{\frac{3}{2}}\pi+in\Delta\theta\biggr).
\end{equation}
Once we take $q_0$ and $q_1$ to infinity, such a wormhole exists for all values of $\tilde{n}$ and hence allows us to sum over all $n$ values. The above sum is convergent when $\Delta\theta$ is real. However, it can fail to converge if $\Delta\theta$ is allowed to take arbitrary complex values. The imaginary distance bound (IDB) appears when analytically continuing $\Delta\theta$ to an imaginary value such that the sum remains convergent \cite{DiUbaldo:2026rly}. It is possible iff
\begin{equation}
    \label{eq:IDB_bound}
    |{\rm Im}(\Delta\theta)|< \sqrt{\frac{3}{2}}\pi\equiv 2\,\tau_{\rm IDB}.
\end{equation}
$\Delta\theta$ represents the boundary value of the axion. The right-hand side of the inequality above is precisely the IDB for a flat space full wormhole \cite{Maldacena:2026jqd, DiUbaldo:2026rly}. As discussed in sec. \ref{sec:Neumann-wormhole} $\tau_{IDB}$ corresponds to a half wormhole geometry where one sets the left boundary of the wormhole to the throat and sends the right boundary to the asymptotic.

It is important to contrast our work with \cite{Maldacena:2026jqd,DiUbaldo:2026rly}. While the authors in \cite{Maldacena:2026jqd,DiUbaldo:2026rly} work directly in the asymptotic limit and assume the saddle contributes to the path integral. However, in this work, we formulate the problem to assess the relevance of the saddles. After figuring out the relevant and irrelevant saddles, we then take the asymptotic analysis to figure out which saddle corresponds to the action of an `` imaginary wormhole ". Our conclusion is that the irrelevant saddles correspond to the imaginary wormholes. This conclusion depends on our choice of integration contour. Although we consider only a half-line contour $N_c\in (0,\infty)$, by extending to all the way to $-\infty$, while going either above or below the singularity, these saddles remain irrelevant, see figure \ref{fig:PL_plot_euclidean_dirichlet}. There is no Lorentzian contour that makes the saddles relevant in the path integral \footnote{A qualitatively similar conclusion was reached in \cite{Held:2026huj}, albeit in a different setting, where the wormhole saddles were found to be irrelevant for imaginary axion values. The saddle $N_{--}$ is the dominant for sufficiently large $q_0$ and $q_1$. This related scenario arises in quantum cosmology \cite{Feldbrugge:2017kzv}, where the dominant saddle is also found to be irrelevant by the Lorentzian contour. }. We already emphasized that the integration contour should be part of the wormhole contribution. It would therefore be worthwhile to investigate whether alternative contours render these saddles relevant. For definiteness, however, we restrict the present analysis to the positive real lapse contour.

Now we will move to the scalar side of the duality and try to recover the imaginary bound directly from the saddles of the scalar axion.

%%%%%%%%%%%%%%%%%%%%%%%%%%%%%%%%%%%%%%%%%%%%%%%

\subsection{Scalar side of the duality}
\label{sec:Dirichlet_scalar_axion_side}

\begin{figure}
  \centering
  \includegraphics[width=0.65\linewidth]{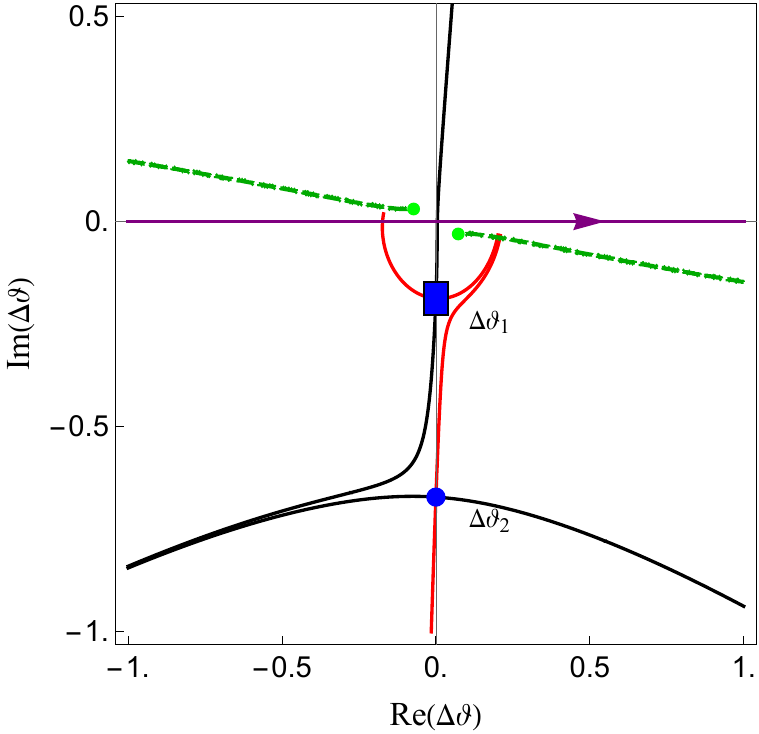}
  \caption{PL plot in the
complex $\Delta\vartheta$-plane for $q_0=1.03$, $q_1=1.1$, $\tilde n=1$, $n=48$.
The blue square/circle denotes the two saddles $\Delta\vartheta_1$
and $\Delta\vartheta_2$,  $(\epsilon_1,\epsilon_2)=(+1,-1)$ and $(-1,-1)$. Clearly, the $(-1,-1)$ saddle is not relevant according to the PL. This is the saddle that corresponds to the IDB in the asymptotically flat limit, as explained in the main text. The green curve is the branch cuts.
The purple line denotes the integration contour
$\tilde{\mathcal{C}}:\Delta\vartheta\in(-\infty,\infty)$.}
  \label{fig:thetaplane}
\end{figure}

Let us now turn to the scalar side of the duality. The path integral for gravity coupled to the massless scalar axion as given in eq. (\ref{eq:path_integral_axion}), reads as ($f_\theta=1,8\pi G=1$)
\begin{equation}
    \label{eq:fixed_charge_path_int}
    \begin{split}
    &\psi[q_0,q_1;\Delta\theta]=\int_0^\infty dN_c\int \mathcal{D}q \int\mathcal{D}\theta\exp[iS_{\rm axion}[\theta,q,N_c]\\
    & S_{\rm axion}[\theta,q,N_c]=2\pi^2\int_0^1\biggl[3N_c-\frac{3\dot{q}^2}{4N_c}+\frac{q^2\dot{\theta}^2}{2 N_c}\biggr],
    \end{split}
\end{equation}
where we added appropriate surface terms $S_{\rm bd}$.
The above action has the shift symmetry $\theta\rightarrow\theta+c$, $c$ is constant. $\Delta\theta$ is the difference of the boundary values of the axion. We choose the axion to vanish at the initial hypersurface and hence $\Delta\theta $ is the value of the axion on the final hypersurface. This is in contrast to \cite{Maldacena:2026jqd}, which considered the axion to be vanishing at the throat. Owing to the shift symmetry, both circumstances describe the same physical situation. The equation of motion for $q$ and $\theta$ are given by
\begin{equation}
    \label{eq:eom_q_theta}
   \ddot{q}+q\dot{\vartheta}^2=0,\quad\frac{d}{dt}(q^2\dot{\vartheta})=0,
\end{equation}
where we have introduced a rescaled variable $\vartheta=\sqrt{2/3}\theta$ \footnote{The normalization of $\sqrt{2/3}$ in the axion is explained from $S^1$-compactification of the $5D$ metric \cite{Maldacena:2026jqd}}. As shown in the appendix \ref{sec:stres_tensor_equality} under the duality transformation, the above equations become identical to the eq. (\ref{eq:on-shell}). The minisuperspace metric with ($q,\vartheta$) coordinate now becomes $ds^2_{\rm mini}=-dq^2+q^2d\vartheta^2$. The solution of eq. (\ref{eq:eom_q_theta}) takes the following form
\begin{equation}
    \label{eq:solution_qtheta_eom}
    \begin{split}
        & q_\pm(t)=\sqrt{q_0^2(1-t)^2+q_1^2 t^2\pm2 q_0q_1\cosh(\Delta\vartheta)t(1-t)},\\
        &\vartheta_\pm(t)=\frac{1}{2}\ln\biggl(\frac{q_0(1-t)\pm q_1 e^{\Delta\vartheta}t}{q_0(1-t)\pm q_1 e^{-\Delta\vartheta}t}\biggr).
    \end{split}
\end{equation}
As in the three-form side of the duality, the solutions separate into two branches, with only the $+$ branch corresponding to a physically admissible solution, which we consider in the following. Note that the above solutions are independent of $N_c$. Plugging the above solutions into the action eq. (\ref{eq:fixed_charge_path_int}), we are left with an integration over $N_c$, which can be evaluated as follows
\begin{equation}
    \label{eq:n-integration}
    \begin{split}
\psi[q_0,q_1;\Delta\theta]&\simeq\int_0^{\infty} dN_c\,\exp\biggl[6\pi^2i\biggl(N_c-\frac{b}{4N_c}\biggr)\biggr]=i\sqrt{b-i\ep}\,\,K_1(6\pi^2\sqrt{b-i\ep}),
    \end{split}
\end{equation}
where, $b=q_0^2+q_1^2-2q_0q_1\cosh(\Delta\vartheta)$. We have chosen an $i\epsilon$ prescription to ensure convergence. %lift the branch points away from the real axis.
In the semiclassical limit $\hbar\to 0$, utilizing the approximation $K_{1}(z)\sim \sqrt{\pi/2}\,e^{-z}/\sqrt{z}$, we get $\psi[q_0,q_1;\Delta\theta]\simeq\exp(-6\pi^2 \sqrt{b})$. $\Delta\theta$ is the part of the boundary condition which is an input to the theory and can't be fixed a priori. Now comes the crucial point. To arrive at the other side of the duality, we need to take the integral transform of the fixed axion path integral eq. (\ref{eq:fourier_transform}), yielding
\begin{equation}
    \label{eq:scalar-axion-ft}
    Z_n[q_0,q_1]=\int_{-\infty}^{+\infty} d(\Delta\theta)\,e^{in\Delta\theta}\psi[q_0,q_1;\Delta\theta]=\int_{-\infty}^{+\infty} d(\Delta\vartheta)\,\exp[i\Phi(\Delta\vartheta)],
\end{equation}
where $\Phi(\Delta\vartheta)=6\pi^2(\tilde{n}\Delta\vartheta+i\sqrt{b-i\ep})$. The presence of $\epsilon$ lifts the branch cuts away from the real axis\footnote{Our convention of cut is: $\sqrt{z}$ has a cut when ${\rm Re}(z)<0$ and ${\rm Im}(z)=0$.}. Observe that $\Delta\vartheta\rightarrow-\Delta\vartheta$ symmetry present in the fixed $\Delta\vartheta$ path integral is broken by the Fourier transform. However, as a consequence of that, the saddles also inherit the same symmetry.
The saddles of $\Phi(\Delta\vartheta)$ are
\begin{equation}
    \label{eq:saddles-fixed-del-theta}
(\Delta\vartheta)_{\ep_1,\ep_2}=i\biggl[\ep_1\cos^{-1}\biggl(\frac{\tilde{n}}{q_0}\biggr)+\ep_2\cos^{-1}\biggl(\frac{\tilde{n}}{q_1}\biggr)\biggr].
\end{equation}
Interestingly, these saddle points coincide with those obtained earlier in the $(u)$-plane in Eq. \ref{eq:saddle_u_plane}. This is because the above is the covering space representation of the periodic axion. This correspondence holds only at the critical points: away from the saddles, $\Phi(\Delta\theta)$ and $S[u]$ are distinct off-shell functions with different analytic structures. For instance, $S[u]$ is a meromorphic function, whereas $\Phi(\Delta\theta)$ contains a branch cut. Hence, the structure of Lefschetz thimbles and the PL analysis will differ in two planes.
In the parameter space $\tilde{n}<q_0<q_1$, taking $q_0,q_1\rightarrow\infty$ in the above saddles, we obtain
\begin{equation}
    \label{eq:asymptotic_saddle} (\Delta\vartheta)_{\ep_1,\ep_2}\biggr|_{q_0,q_1\rightarrow\infty}=i\frac{\pi}{2}(\ep_1+\ep_2)+\mathcal{O}\biggl(\frac{\tilde{n}}{q_0},\frac{\tilde{n}}{q_1}\biggr),
\end{equation}
which is purely imaginary. This agrees with the fact that to keep the throat of the wormhole stable, one requires scalars to be imaginary valued \cite{Witten:2026twr}. As also shown in appendix \ref{sec:stres_tensor_equality}, the duality requires the scalar to be imaginary for the real three form flux. In the case when $\ep_1\ep_2=1$, we recover the same bound on the imaginary part of $\Delta\theta$ ($\Delta\theta=\sqrt{3/2}\Delta\vartheta$):
\begin{equation}
    \label{eq:bound_from_scalar}
   \ep_1\ep_2=1,\quad \Rightarrow \quad |{\rm Im}(\Delta\theta)|=\sqrt{\frac{3}{2}}\pi\equiv 2\,\tau_{\rm IDB}
\end{equation}
These configurations are dual to the cross-throat saddles on the scalar side. It is therefore the imaginary-wormhole saddles that determine the bound on the imaginary part of the boundary axion separation. For $\ep_1\ep_2=-1$, however, this separation vanishes in the asymptotic limit, since the two boundaries approach the same right asymptotic region and eventually coincide. The scalar profile in eq. (\ref{eq:solution_qtheta_eom}) at these saddles becomes (up to an irrelevant shift)
\begin{equation}
    \label{eq:scalar_axion_profile}
    \theta\biggr|_{\rm cross-throat}=i\sqrt{\frac{3}{2}}\tilde{\tau},\quad \cos\tilde{\tau}=\biggl(\frac{a_{\rm th}}{a}\biggr)^2.
\end{equation}
The above scalar profile at the saddles coincides with \cite{Maldacena:2026jqd}. Evaluating the on-shell action at the saddles, we find
\begin{equation}
    \label{eq:dirichket-axion action}
    \begin{split}
&\psi\big[q_0,q_1;\Delta\theta\big]\Big|_{\epsilon_1,\epsilon_2}\sim e^{\,I_D},
\quad
I_D\big(\epsilon_1,\epsilon_2\big)=-6\pi^{2}\left|\sqrt{q_0^{2}-\tilde n^{2}}
+\epsilon_1\epsilon_2\sqrt{q_1^{2}-\tilde n^{2}}\,\right| ,
    \end{split}
\end{equation}
Although in the large-$q_0,q_1$ limit $I_D$ diverges (divergence is linear in $q_0$ and $q_1$), after background subtraction it is zero because there is no additional finite part: $I_D^{\rm ren}=0$ \cite{Maldacena:2026jqd}. The vanishing of the Dirichlet action explains why the fixed-$n$ action is linear in $|n|$ via the Legendre transform. As shown in fig. \ref{fig:thetaplane}, according to the Picard-Lefschetz method, the saddles corresponding to the cross-throat geometry ($\ep_1\ep_2=1$) are irrelevant in the path integral. This also aligns with the expectation from duality. Hence, analyzing both sides of the duality using the Picard-Lefschetz, we arrive at the same conclusion that the imaginary wormhole corresponding to the distance bound (IDB) is not relevant in the path integral when the contour of metric integration is defined along the Lorentzian contour $N_c\in(0,\infty)$.

%%%%%%%%%%%%%%%%%%%%%%%%%%%%%%%%%%%%%%%%%%%%%%%%%%
%%%%%%%%%%%%%%%%%%%%%%%%%%%%%%%%%%%%%%%%%%%%%%%%%%

\section{Wormhole saddles with Neumann condition} 
\label{sec:Neumann-wormhole}
In the previous section, we considered the Dirichlet boundary condition on the metric. We noticed that the saddle configurations describe two different types of geometry, namely, same-side and cross-throat. These are two different segments of the full GS wormhole as illustrated in the figure \ref{fig:saddle-geometry}. Utilizing the Picard-Lefschetz analysis, we showed that ``cross-throat" saddles are always irrelevant and do not participate in the path integral as the steepest ascent line from these saddles doesn't intersect the integration contour. Upon taking the large asymptotic limits, the cross-throat saddles eventually become the full GS wormholes connecting to asymptotically flat spaces and also satisfy the imaginary distance bound on the scalar axion. We analyzed both sides of the duality and reached the same conclusion. 

Having found that the saddles are not relevant to the Dirichlet condition, we are prompted to check whether one can make them relevant either by modifying the contour or by changing the boundary condition on the metric. Different boundary conditions induce different off-shell structures, keeping the saddle geometries the same. Concrete realizations of this idea arise in quantum cosmology \cite{Lehners:2023yrj, Ailiga:2023wzl, Ailiga:2024wdx, DiTucci:2019bui,Lavrelashvili:2026jcw}, where changing the boundary conditions modifies the Picard-Lefschetz decomposition and, consequently, the set of relevant saddle points. Having found that there is no simple extension of the Lorentzian contour that can make the saddles relevant, we consider the second alternative and choose the simplest modification by choosing the following Neumann condition on the metric:
\begin{equation}
    \label{eq:neuman_bc}
    {\rm Bd_q(0)}: \,\frac{\dot{q}(0)}{N_c}=\alpha,\quad{\rm Bd_q(1)}: \,q(1)=q_1,
\end{equation}
where we allow any arbitrary complex values of $\alpha$. A second motivation for considering the Neumann condition is its behaviour under metric scaling ($g_{\mu\nu}\rightarrow\lam g_{\mu\nu}$). As the Neumann condition remains invariant under such a scaling transformation, it yields distinctive features in the asymptotic regime. This will play an important role in the discussion of the imaginary distance bound (IDB) and KSW allowability. Since the action is singular at $q=0$, to make $\alpha$ finite, we restrict to the sheet where always \footnote{As we discuss in sec. \ref{sec:ksw}, the choice ${\rm Re}(q(t))<0$ is also at odd with the KSW criterion.} ${\rm Re}(q(t))>0$. This aligns with the same convention taken in Dirichlet in sec. \ref{sec:saddle_and_PL}. Although $\alpha$ is arbitrary, only for a restricted range, one recovers the saddle geometries describing Euclidean wormholes. For generic values of $\alpha$, the boundary condition instead selects a distinct class of complex background geometries. As the duality relating the three-form flux and axion is argued to be independent of any background \cite{Witten:2026twr}. The setup can also be used to explicitly verify the claim using the Picard-Lefschetz method. This is a non-trivial test, as modifying the boundary condition changes the thimble structure, which in turn affects the relevance of the contributing saddles in the path integral and could thus spoil the duality. It's not guaranteed that the dominant saddle in the one-representation maps to the dominant saddle in the dual representation. In this section, we will also address this question. 

%%%%%%%%%%%%%%%%%%%%%%%%%%%%%%%%%%%%%%%%%%%%%%%%%%%

\subsection{3-form side of the duality}
\label{eq:neumann-3-form-side}
To identify the wormhole saddles when the Neumann condition is imposed on the metric, we start from the three-form side of the duality. The action with the appropriate surface term, as in eq. (\ref{eq:path_integral_3form}) is given by
\begin{equation}
    \label{eq:3-form-neumann}
    S^N_{3-{\rm form}}[q,h]=2\pi^2\int_0^1 dt\biggl[3N_c-\frac{3\dot{q}^2}{4N_c}-\frac{h^2N_c}{2 q^2}\biggr]-3\alpha\pi^2 q(0),
\end{equation}
where, we put the subscript ``$N$" to denote Neumann case. The action above admits a well-posed variational principle under the boundary condition in eq. (\ref{eq:neuman_bc}). Unlike the Dirichlet condition, the path integral can't be solved exactly, so we proceed using the saddle point approximation. The equation of motion for $q(t)$, following from the above action, is given in eq. (\ref{eq:on-shell}). Solving it subject to the mixed Neumann-Dirichlet condition, we get
\begin{equation}
\label{eq:q(t)_for_three_form}
\begin{split}
&q(t)=
\sqrt{
\left(\mathbf{r}+\alpha N_{c}t\right)^{2}
-\frac{4\tilde{n}^{2}N_{c}^{2}t^{2}}{\mathbf{r}^{2}}
}.\\
\text{where,}\qquad& \mathbf{r}^4
+2\alpha N_c \mathbf{r}^3
+\left(\alpha^2N_c^2-q_1^2\right)\mathbf{r}^2
-4\tilde{n}^2N_c^2
=0.
\end{split}
\end{equation}
The on-shell equation admits four distinct branches, corresponding to the four roots of the quartic equation. However, not all of them are physical branches. To identify the physical branch, we examine the behavior of the solutions as $(N_c\to0)$. In this limit, the four branches approach $\mathbf r\rightarrow 0,0,+q_1,-q_1$.
Requiring $({\rm Re}(\mathbf r)>0)$ uniquely selects the physical branch for which $\mathbf r\to q_1.$
Plugging the above on-shell equation into the action in eq. (\ref{eq:3-form-neumann}), we obtain the following off-shell action for lapse integration
\begin{equation}
\label{eq:on-shell_action_neumann}
S_N(N_c)=
3\pi^2
\left[
2N_c
-\alpha \mathbf{r}
-\frac{\alpha^2N_c}{2}
+\frac{2\tilde{n}^2N_c}{\mathbf{r}^2}
-2\tilde{n}\,
\tanh^{-1}\!\left(
\frac{2\tilde{n}N_c}
{\mathbf{r}\left(\mathbf{r}+\alpha N_c\right)}
\right)
\right].
\end{equation}
For the physical branch, $N_c=0$ is a regular point and not the essential singularity of the action. As each of the $\mathbf{r}$ branches depends on $N_c$, the saddles of the above action are found utilizing the chain rule $\partial S_N/\partial N_c+(\partial S_N/\partial\mathbf{r})(\partial\mathbf{r}/\partial N_c)=0$, yielding
\begin{equation}
\label{eq:saddle_expression}
N_{\rho,\eta}=\frac{1}{2}
\left[
\rho
{\frac{\tilde{n}\alpha}{\sqrt{4+\alpha^{2}}}}
+
\eta
\sqrt{\tilde{n}^{2}-q_{1}^{2}}
\right],
\qquad
\rho,\eta=\pm1.
\end{equation}
For a generic nonzero value of $(\alpha)$, the lapse action admits four saddle points. The corresponding geometries are obtained by substituting these saddle-point values into the metric ansatz in Eq.~(\ref{eq:metric}). As shown in appendix \ref{app:GS_from_Neumann}, the resulting metric takes the identical form as the Giddings-Strominger wormhole geometry given in Eq.~(\ref{eq:Giddings-stominger-wormholes}). Although $\alpha$, $\rho$ and $\eta$ drop out of the line element, they determine the location, orientation and segment of the wormhole. Evaluating the on-shell action at these saddles, we get
\begin{equation}
\label{eq:action_at_saddle_form_Side}
S_N\!\left(N_{\rho,\eta}\right)
=
6\pi^{2}
\biggl[
\eta\biggl\{\sqrt{\tilde{n}^{2}-q_{1}^{2}}
-\tilde{n}\cosh^{-1}\biggl(\frac{\tilde{n}}{q_1}\biggr)\biggr\}+
\tilde{n}\ln\!\biggl(
\frac{
\sqrt{4+\alpha^{2}}-\rho\alpha
}{
2
}
\biggr)
\biggr].
\end{equation}
We will need the above action at the saddles in analyzing the scalar axion side of the duality. Observe that the above action at the saddle splits into a part depending on the final parameter ($q_1$) and initial parameter ($\alpha$), similar to the Dirichlet condition.

%%%%%%%%%%%%%%%%%%%%%%%%%%%%%%%%%%%%%%%%%%%%%%%

\subsubsection{Bound On $\alpha$}
\label{eq:bound_alpha_}
We have shown that, for arbitrary $\alpha$, the saddle-geometries take the same metric form as the GS wormhole metric in eq. (\ref{eq:Giddings-stominger-wormholes}). However, for generic $\alpha$ with $q_1>\tilde{n}$, it will be a complex continuation of the Euclidean GS wormholes. It is therefore important to identify the parameter space that admits real Euclidean wormholes, as well as imaginary wormholes compatible with the imaginary distance bound. This is the purpose of the present section.

First, we note that to obtain geometries that are real and Euclidean, one should have saddles that are purely imaginary. From the expression of the saddles in eq. (\ref{eq:saddle_expression}), we arrive at the conclusion that $\alpha$ has to be purely imaginary and its absolute value must lie between zero and two. Furthermore, one arrives at the same conclusion by comparing the Neumann saddles to the Dirichlet saddles as in eq. (\ref{sec:saddle_dirichlet}). Hence, we arrive at the following bound on the $\alpha$ to get a saddle geometry which is a segment of Euclidean GS wormhole
\begin{equation}
    \label{eq:bound_on_alpha}
  \alpha=i\ep|\alpha|,\qquad  0\leq |\alpha|\leq 2,\qquad \ep=\pm 1.
\end{equation}
\begin{figure}
    \centering
    \includegraphics[width=0.65\linewidth]{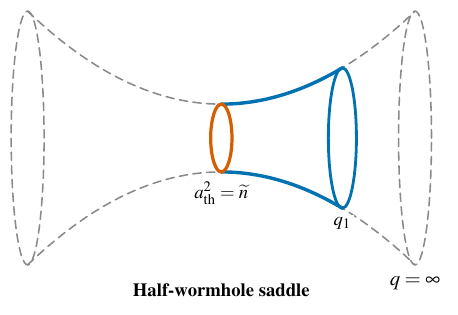}
    \caption{The saddle geometry realized by $\alpha=0$. The grey dashed background is the full wormhole profile, while the blue portion describes the saddle segment. When we take $q_1$ into the right asymptotic region, we obtain the half-wormhole with a boundary only on the left (Half WH). Such geometries give rise to the $\tau_{\rm IDB}$.}
    \label{fig:half-wormhole}
\end{figure}
$|\alpha|=0$ corresponds to a half wormhole starting from the throat at $a_{\rm th}^2=\tilde{n}$ to the right asymptotic boundary, see figure \ref{fig:half-wormhole}. This corresponds to a geometry with a finite boundary on the left. For non-zero $|\alpha|$, the left boundary of the wormhole moves either to the left or to the right depending on the sign of $\rho,\epsilon,\eta$. The branch with $\rho\eta\ep=-1$ corresponds to the same-side geometry, whereas $\rho\eta\ep=+1$ yields the cross-throat geometry, as illustrated in fig. \ref{fig:saddle-geometry}. Now the sign of $\rho$ is related to the sign of $\mathbf{r}$, which must be positive ($\mathbf{r}>0$), and hence, we choose $\rho=+1$. Hence, we arrive at the following conclusions
\begin{equation}
    \label{eq:cross-throat_neumann}
    \begin{split}
    &\ep\eta=+1\quad \Longrightarrow \text{Cross-throat saddles},\quad \ep\eta=-1\quad \Longrightarrow \text{Same-side saddles}.
    \end{split}
\end{equation}

In the limiting case $|\alpha|\to 2$, the cross-throat branch describes an imaginary wormhole. The bound on $\alpha$ in eq. (\ref{eq:bound_on_alpha}) plays a key role in the following discussion of the imaginary distance bound. Generic complex $\alpha$ instead gives the corresponding complexified GS geometry, see table \ref{tab:dirichlet-neuman-saddle-relevance}. 

\begin{table}[hbtp]
\centering
\begin{tabular}{|c|c|}
\hline
    value of $\alpha$ & initial size $q(0)$
    \\ 
    \hline 
     $\alpha\in \mathbb{R}-\{0\}$ & $q(0)<\tilde{n}$ \\
    \hline
    $\alpha=\pm i|\alpha|,0<|\alpha|<2$ & $q(0)>\tilde{n}$ \\
    \hline
    $\alpha=\pm i|\alpha|,|\alpha|>2$ & $q(0)$ complex \\
    \hline
    $\alpha=0$& $q(0)=\tilde{n}$\\
    \hline
    $\alpha=x+iy,x\neq0,y\neq 0$& $q(0)$ complex\\
    \hline
\end{tabular}
\caption{Nature of saddles for different values of $\alpha$. This connects the Dirichlet saddles with Neumann saddles. In the regime where $q(0)>\tilde{n}$, the geometry describes an Euclidean wormhole.}
\label{tab:dirichlet-neuman-saddle-relevance}
\end{table}

%%%%%%%%%%%%%%%%%%%%%%%%%%%%%%%%%%%%%%%%%%%%%%%%%%%%

\subsubsection{Asymptotic behaviour and the imaginary wormholes}
\label{eq:neumann-imaginary-wormhole}

In this section, we examine the asymptotic behavior of the wormhole geometries in the regime $0\leq |\alpha|\leq 2$ and derive the corresponding distance bound. The analysis closely parallels that of the Dirichlet case, with one important distinction: owing to the invariance of the Neumann boundary condition under metric rescaling, no divergent contribution arises in the asymptotic expansion. In the Dirichlet end, however, one encounters the usual divergence. Taking the large-$q_1$ limit of eq. (\ref{eq:action_at_saddle_form_Side}), we obtain
\begin{equation}
    \label{eq:asymptotic_express_neumann}
    iS_N(N_{\rho,\eta})\biggr|_{q_1\rightarrow R_2}=6\pi^2
\left[
-\eta R_2
+i\tilde{n}\,
\ln\!\left(
-i\eta\,
\frac{\sqrt{4+\alpha^2}-\rho\alpha}{2}
\right)
+\mathcal{O}\!\left(R_2^{-1}\right)
\right],
\end{equation}
where $R_2$ is the asymptotic cut-off. Subtracting the divergent part, which is again linear in $R_2$, we get the following renormalized action at the saddles,
\begin{equation}
    \label{eq:renormalized_neumann}
    iS_N(N_{\rho,\eta})\biggr|_{\rm ren}=i6\pi^2\tilde{n}\ln\!\left(
-i\eta\,
\frac{\sqrt{4+\alpha^2}-\rho\alpha}{2}
\right).
\end{equation}
Substituting $\alpha=2i\ep,\ep=\pm 1$ in the above expression, we get for the choice $\ep\rho\eta=1$,
\begin{equation}
    \label{eq:condition_sign_alpha}
    \ep \rho\eta=1\quad \Longrightarrow\quad iS_N(N_{\rho,\eta})\biggr|_{\rm ren}=-6\pi^3\tilde{n}=-\sqrt{\frac{3}{2}}|n|\pi,
\end{equation}
where, we have utilized $\tilde{n}=|n|/2\pi^2\sqrt{6}$.
For $\ep\eta\rho=-1$, the renormalized action vanishes. By contrast, the branch $ \ep \rho\eta=1$ corresponding to the cross-throat geometry, yields precisely the same Dirichlet exponent in eq. (\ref{eq:mmm_saddle}) and therefore reproduces the same imaginary distance bound. Thus, an imaginary wormhole geometry is obtained by imposing the following Neumann boundary condition \footnote{Interestingly, this boundary condition coincides with the Euclidean-momentum condition employed in Lorentzian implementations of the no-boundary proposal \cite{Lehners:2023yrj, Ailiga:2023wzl, Ailiga:2024wdx, DiTucci:2019bui}. See \cite{Lavrelashvili:2026jcw}, where the authors also study the Neumann condition on metric for wormholes. } 
\begin{equation}
\label{eq:boundary_condition}
\text{Imaginary wormhole}
\quad\Longrightarrow\quad
\frac{\dot q}{2N_c}=\pm i .
\end{equation}

Let us now consider the sum over fixed flux $n$  wormholes for $\alpha$ away from the imaginary bound. The geometry interpolates between the half wormhole ($|\alpha|=0$) and the complete wormhole. This behavior contrasts with the Dirichlet case, where the sum over $n$ at fixed $\alpha$ was not possible. At fixed $q_0$, the Euclidean regime $\tilde{n}<q_0<q_1$ fails once $|n|>2\pi^2q_0/\sqrt{6}$. The saddles turn real, and the geometry leaves the Euclidean wormhole family. With fixed $\alpha$, $q_0$ scales up with $n$ automatically, so no sector is ever expelled through the $t=0$ end. This is a consequence of the scaling invariance of the Neumann condition, and allows one to sum over $n$ for $\alpha$ away from its critical value. The sum over $n$, as defined in eq. (\ref{eq:n_sum_identity}) takes the following form (performing the background subtraction at $t=1$ end and $\ep=\eta=-1$)\footnote{$\ln(z)$ has a branch cut along the negative real axis. According to the convention $\ln(-1\pm i\ep)=\pm i\pi$ \cite{DLMF}}
\begin{equation}
\label{eq:bound_on_theta_alpha}
\begin{split}
&Z^\alpha[\Delta\theta]=\sum_{n\in \mathbb{Z}}e^{i\Delta\theta n}Z_n^\alpha=\sum_{n\in\mathbb{Z}}\exp\biggl[in\Delta\theta+i\sqrt{\frac{3}{2}}|n|\ln\biggl(\frac{i\sqrt{4-|\alpha|^2}-|\alpha|}{2}\biggr)\biggr] \\
& \text{convergence requires,}\,\,  \left|\operatorname{Im}\Delta\theta\right|+\sqrt{\frac{3}{2}}\arccos\left(\frac{|\alpha|}{2}\right)<
\sqrt{\frac{3}{2}}\pi,
\end{split}
\end{equation}
where $\Delta\theta$ is the difference between the boundary values of the axion.
$Z_n^\alpha$ in the above equation is the exponentiation of the renormalized lapse action as given in eq. (\ref{eq:renormalized_neumann}). The above equation generalizes the bound for the imaginary value of the boundary axion for generic $\alpha$, which in the special case of $|\alpha|=0,2$ reduces to the old bound for half and full wormholes, respectively. The above bound in eq. (\ref{eq:bound_on_theta_alpha}) is one of the main results of our paper, which we revisit in sec. \ref{sec:ksw} while discussing the KSW criterion. We independently derive this bound from the KSW allowability criterion. The explicit $\alpha$- dependence makes its connection to the imaginary distance bound transparent—a feature absent under Dirichlet boundary conditions. At the end of the computation, we put $|\alpha|=2$ to recover the bound associated with a complete wormhole profile. \\

\textbf{Half Wormhole:} Let us consider the case of a half-wormhole, which corresponds to the case $|\alpha|=0$. The wormhole geometry is shown in fig \ref{fig:half-wormhole}. Upon taking the large-$q_1$ limit and performing asymptotic analysis, this yields a geometry with a single boundary at the throat, while the other end approaches the right-asymptotic region. To determine the imaginary distance bound associated with a half-wormhole, we plug $|\alpha|=0$ into the inequality \ref{eq:bound_on_theta_alpha}, yielding
\begin{equation}
    \label{eq:half_wormhole_bound}
    |{\rm Im}(\Delta\theta)|_{\rm Half-WH}<\frac{1}{2}\sqrt{\frac{3}{2}}\pi\equiv \tau_{\rm IDB}.
\end{equation}
Hence, the imaginary distance bound for a half-wormhole is half that of a full wormhole profile \cite{Maldacena:2026jqd}.

\begin{figure}[htbp]
  \centering  \subfigure{\includegraphics[width=0.65\linewidth]{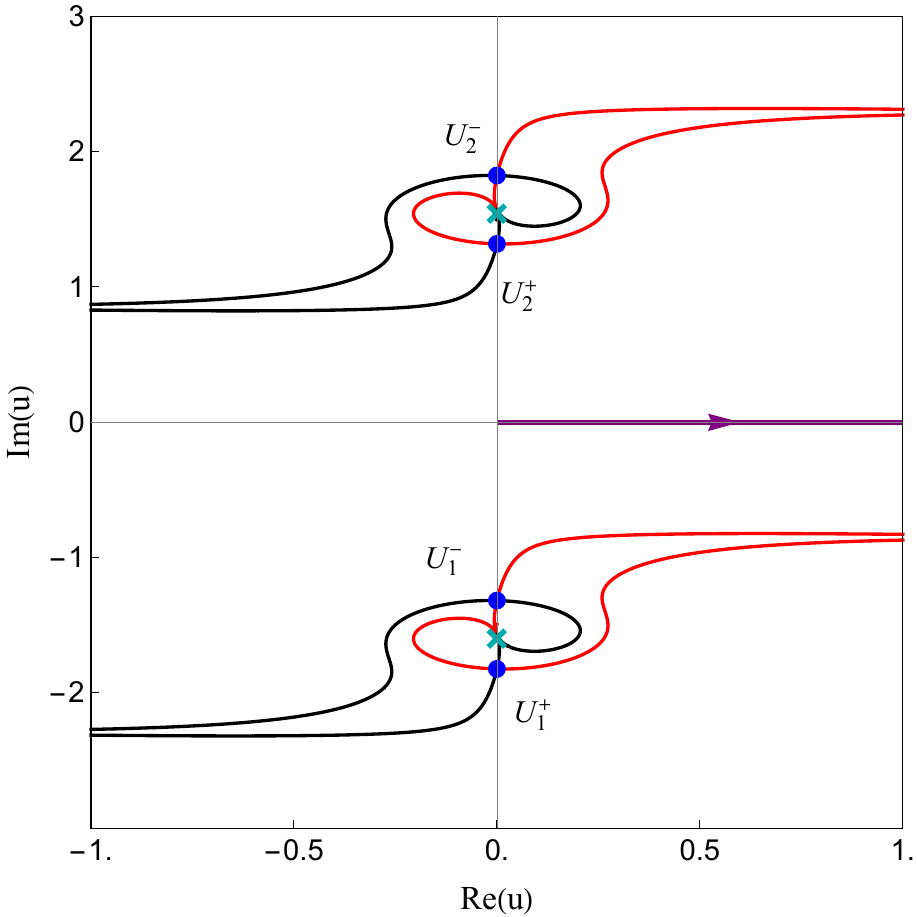}}
  \caption{PL plot for $\alpha=0,$ $ q_1=4,$ $ \tilde{n}=1,$$ n=48$ in the covering $u$-plane. The geometry describes a half-wormhole saddle (fig. \ref{fig:half-wormhole}). $U_1^\pm,U_2^\pm$ are the images of $\eta=\pm$ saddles of $N_c$-plane. According to PL, no saddle is relevant in both the figures.}
  \label{fig:threeform alpha zero}
  \end{figure}

%%%%%%%%%%%%%%%%%%%%%%%%%%%%%%%%%%%%%%%%%%%%%%%%%%%%%

\subsubsection{PL relevance of wormhole saddles}
\label{sec:relevance_neumann_saddle}

\begin{figure}
    \centering
    \includegraphics[width=1\linewidth]{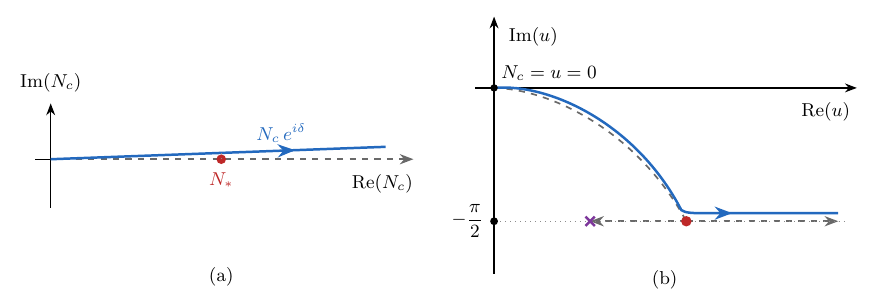}
    \caption{For $\beta<\beta_{\rm crit}$, the physical lift of the real $N_c$ contour (dashed curve) in the $u$-plane meets the geometries with $q(t_*)=0$, as shown by horizontal dashed line. To avoid it, we slightly rotate the real $N_c$ contour ensuring convergence, $N_c\,e^{i\delta},0<\delta\ll 1$. As a result the contour curves around the ramification point $N_*$ (red dot) and doesn't hit. In solid blue we depict (a) contour in $N_c$ plane and (b) the physical lift of the contour in the $u$-plane.}
    \label{fig:physical_lift_rotating contour}
\end{figure}

In this section, we employ Picard–Lefschetz theory to determine the relevance of the wormhole saddles. We are interested in seeing if the cross-throat saddle corresponding to the imaginary wormhole is relevant in the path integral. As in the Dirichlet analysis, it is convenient to lift the problem to a covering plane that incorporates all branches of $\mathbf{r}$ while eliminating the inverse functions in the action and their associated branch cuts. We define the covering coordinate $u$ by
\begin{equation}
    \label{eq:covering_map_neumann}
    \mathbf{r}(u)=
\frac{2\tilde{n} q_1\cosh u}
     {2\tilde{n}+\alpha q_1\sinh u},
\qquad
N_c(u)=
\frac{q_1^{2}\sinh u\cosh u}
     {2\tilde{n}+\alpha q_1\sinh u},\qquad{\rm Re}(\mathbf{r})>0. 
\end{equation}
The above transformation satisfies the quartic equations in \ref{eq:q(t)_for_three_form}, which corresponds to four possible branches. 
Upon transforming to the $u$-plane, the integration over $N_c$ as in eq. (\ref{eq:u-plaane integral}) becomes
\begin{equation}
     \label{eq:u-plaane integral_neumann}
     \begin{split}
    Z_n[\alpha,q_1]=& \int_{\mathcal{C}_u} du \,\mathcal{A}_N(u)\,\exp(iS_N[u])\\
  \text{where},\qquad   \frac{S_N[u]}{6\pi^2}=
&\frac{
\left\{
\tilde{n}^{\,2}
+q_1^2\left(
\cosh^2u-\frac{\alpha^2}{4}
\right)
\right\}\sinh u
-\alpha\tilde{n} q_1
}{
\cosh u\left(
2\tilde{n}+\alpha q_1\sinh u
\right)
}
-\tilde{n} u.
\end{split}
\end{equation}
Clearly $S_N[u]$ is a meromorphic function of $u$ with no branch cut. $\mathcal{A}_N(u)$ vanishes at the ramification points where the covering map degenerates. Each saddle $N_{\rho\eta}$ in the $N_c$-plane has an image in the $u$-plane as $U_{\rho\eta}$. The contour $\mathcal{C}_u$ is the pull-back of the real contour $N_c\in(0,\infty)$ into the $u$-plane via eq. (\ref{eq:covering_map_neumann}). The condition ${\rm Re}(\mathbf{r})>0$ enforces that it is a physical lift of the Lorentzian contour in the covering plane. This is possible only if $u=0$ at $N_c=0$. So the contour should be connected to the origin. Such a criterion, however, can fail if it hits a ramification point and touches the branch which corresponds to the singular histories $q(t_*)=0$ for $t_*\in(0,1)$. For this section, we will stick to the Euclidean regime, where $\alpha=i\ep|\alpha|$. As shown in appendix \ref{sec:physical_lift} for parameters satisfying $\beta\equiv |\alpha|q_1/2\tilde{n}<\beta_{\rm crit}$, $\beta_{\rm crit}=4\sqrt{6}/9$ the physical lift hits the singular histories. This can be remedied by slightly rotating the $N_c$ contour into the complex plane as shown in fig. \ref{fig:physical_lift_rotating contour}. This determines the contour in the $u$-plane where PL is performed. 

To determine the direction of contour near $u=0$, we invert the relation in eq. (\ref{eq:covering_map_neumann}) and obtain for small $N_c$ 
\begin{equation}
\label{eq:invert_contour}
    u_\alpha(N_c)=
    \frac{2\tilde{n}}{q_1^2}\,N_c+
    \frac{2\alpha \tilde{n}}{q_1^3}\,N_c^2+\mathcal{O}(N_c^3),\quad N_c \quad \text{is real}.
\end{equation}
From the above equation, it is easy to convince that for imaginary $\alpha$ with $\ep=+1$, it bends in the upper half plane, $\ep=-1$, it bends in the lower half plane. The analytic equation for contour in the complex $u=u_x+iu_y$-plane, according to the eq. (\ref{eq:covering_map_neumann}), becomes for $\beta<\beta_{\rm crit}$
\begin{equation}
    \label{eq:contour}
    \begin{split}
    \mathcal{C}_u=\{(u_x,u_y):\delta R_\ep (u_x,u_y)-I_\ep(u_x,u_y)=0 \quad\&\quad R_\ep(u_x,u_y)+\delta I_\ep(u_x,u_y)>0\}.
    \end{split}
\end{equation}
In the $N_c$ plane the above contour corresponds to $N_c\rightarrow N_c e^{i\delta}, N_c>0,0<\delta\ll 1$, see fig \ref{fig:physical_lift_rotating contour}.
For $\beta>\beta_{\rm crit}$, one can safely put $\delta=0$. The expressions for $R_\ep$ and $I_\ep$ are (see appendix \ref{sec:physical_lift} for details)
\begin{equation}
\label{R_and_I_main}
    \begin{split}
&R_\ep=2\tilde{n}\sinh(u_x)\cosh(u_x)\cos(2u_y)+\ep|\alpha|q_1\sinh(u_x)\sin(u_y)(\sinh^2 u_x+\sin^2u_y)\\
        &I_\ep=\cos(u_y)[2\tilde{n}\cosh(2u_x)\sin u_y-\epsilon |\alpha| q_1\cosh u_x(\sinh^2 u_x+\sin^2u_y)].
    \end{split}
\end{equation}

\begin{figure}[htbp]
\centering
\subfigure[\,\,$0<|\alpha|<|\alpha|_c$ ]
{
\includegraphics[width=0.473\textwidth]{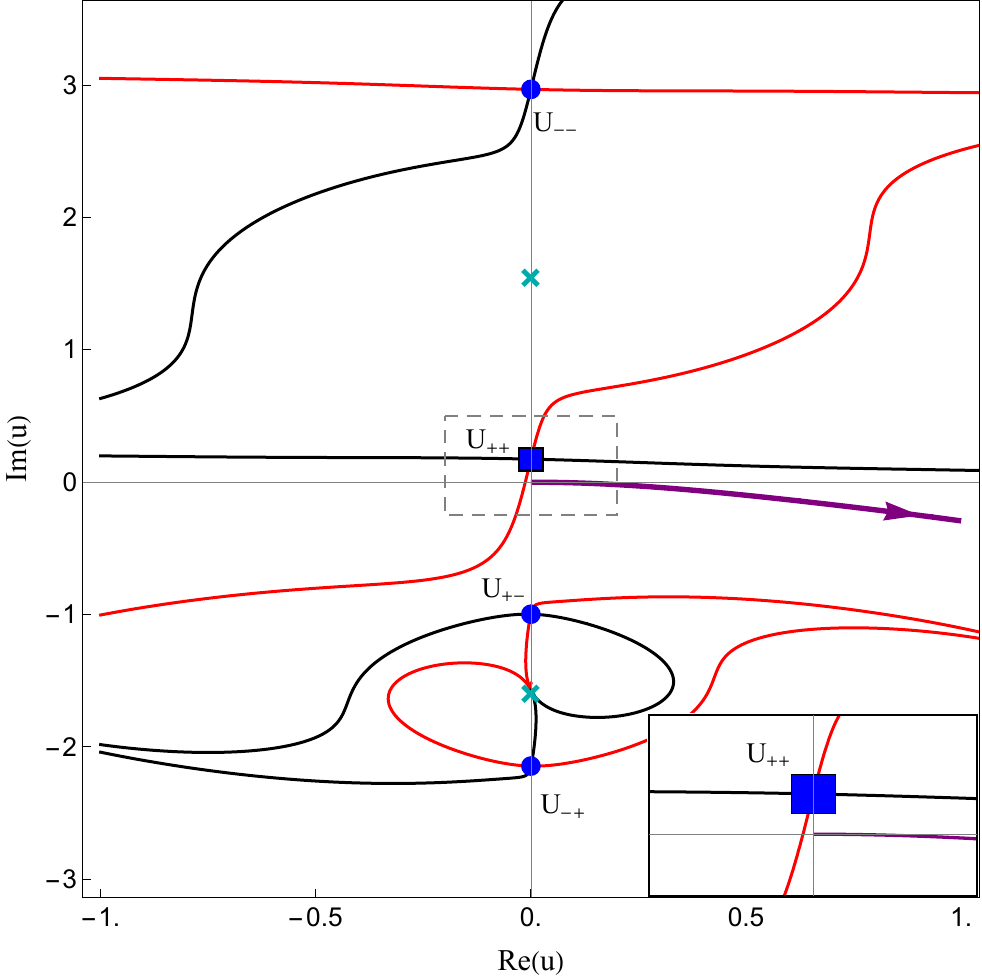}
}
\subfigure[\,\,$|\alpha|_c<|\alpha|<2$]
{
\includegraphics[width=0.473\textwidth]{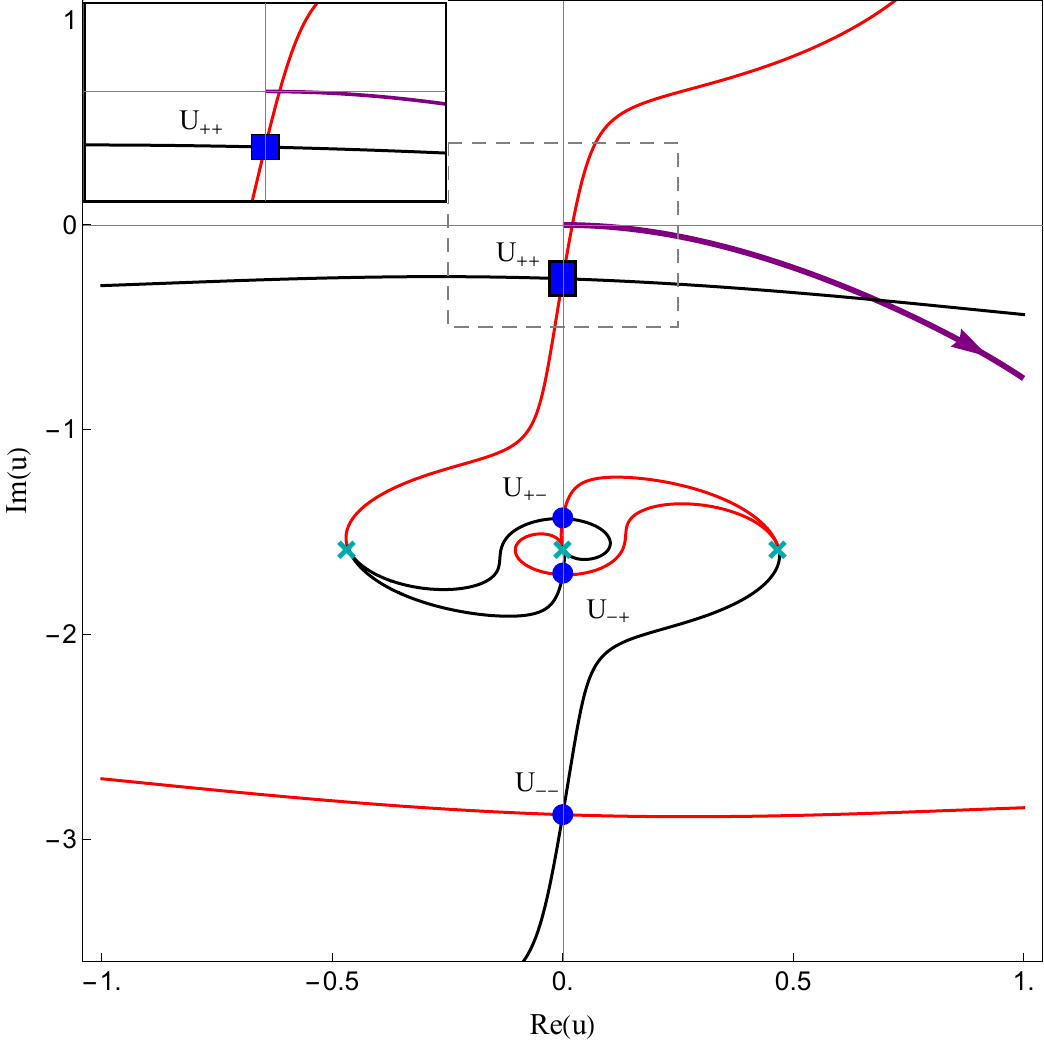}
}
%{
%\includegraphics[width=0.47\textwidth]{alpha1_Threeform.pdf}
%}
\caption{PL plot for wormhole solutions in the range $0<|\alpha|<2$. For $0<|\alpha|<|\alpha|_c$, no saddle is relevant. However for $|\alpha|_c<|\alpha|<2$, one saddle is relevant ($\rho=\eta=1$). We choose $\ep=-1$. In left $ |\alpha|=0.8,$ $q_1=6,$ $\tilde{n}=5$, $ n=242$ and $ |\alpha|=1.5,$ $q_1=6,$ $\tilde{n}=5,n=242$ in right figure. This corresponds to the regime $\beta<\beta_{\rm crit}$. Clearly the cross-throat saddle $U_{+-}$ is irrelevant according to the PL.}
\label{fig:three-form-neumann}
\end{figure}

For $|\alpha|=0$, it moves straight, and the contour becomes a real line $u\in(0,\infty)$. As shown in \ref{fig:threeform alpha zero}, PL analysis reveals that the saddle is not relevant to the path integral.  To perform the Picard-Lefschetz analysis for non-zero $\alpha$, it is useful to define a critical value $|\alpha|_c=2\sqrt{1-\tilde{n}^2/q_1^2}$. The significance of this point is that for the choice $\ep\rho\eta=-1$, the saddle in eq. (\ref{eq:saddle_expression}) reaches the endpoint of the integration contour at $N_c=u=0$ and hence becomes relevant. Choosing $\rho=+1,\ep=-1$, we find that the relevant saddles are %for $\ep\eta=-1$, the saddles become relevant as
\begin{equation}
    \label{eq:relevance}
   0\leq |\alpha|<|\alpha|_c: \,\, \text{No saddle is relevant},\quad |\alpha|_c\leq|\alpha|<2:\,\, U_{++}\,\, \text{is relevant}.
\end{equation}
This is numerically illustrated in fig. \ref{fig:three-form-neumann}. The saddle $U_{++}$ is the image of the $N_{++}$ saddle and belongs to the same side branch $\ep\eta=-1$. There is no saddle that is relevant for $\ep\eta=1$ corresponding to cross-throat geometry. This is also evident from the plot, as the saddle denoted as $U_{+-}$ is irrelevant, which corresponds to the cross-throat geometry for $\ep=-1,\eta=-1$. In the asymptotic limit, the cross-throat saddle approaches the complete Giddings-Strominger wormhole, which can be obtained by glueing two half-wormhole geometries at the throat. Combining this identification with the Picard-Lefschetz analysis, we conclude that both the half-wormhole saddle and the saddle corresponding to the imaginary-wormhole have vanishing intersection number for Neumann boundary conditions and therefore do not contribute to the Lorentzian path integral. 

Now we will turn to the scalar side of the duality and analyze the Picard-Lefschetz relevance of the saddles, yielding the imaginary distance bound.

%We observe that the $\ep\eta=-1$ saddle is always relevant.
%There is always $N_*$ along the real line where $q(t_*)$ vanishes, making the path integral ill-defined.
%$\mathcal{C}=\mathbb{R}_+ e^{i\delta},\,0<\delta\ll 1$.\\

%PL is defined in the physical lift of the Lorentzian contour in the covering $u$-coordinate.

%%%%%%%%%%%%%%%%%%%%%%%%%%%%%%%%%%%%%%%%%%%%%%%%%
\subsection{Axion scalar side of the duality}
\label{eq:axion_neumann_duality}

After analyzing the saddles and thimble structure in the 3-form side of the duality, we now proceed to analyze the same in the scalar axion side. We aim to derive the distance bound as in eq. (\ref{eq:bound_on_theta_alpha}) directly from the saddles of the scalar axion. For the boundary conditions as mentioned in the eq. (\ref{eq:bc_axion}) and eq. (\ref{eq:neuman_bc}) for the axion and metric, respectively, the gravity action coupled to the axion as mentioned in eq. (\ref{eq:path_integral_axion}) evaluates to
\begin{equation}
    \label{eq:axion-neumann}
    S_{\rm axion}[q,\theta]=2\pi^2\int_0^1 dt\biggl[3N_c-\frac{3\dot{q}^2}{4N_c}+\frac{q^2\dot{\theta}^2}{2 N_c}\biggr]-3\alpha\pi^2 q(0).
\end{equation}
The above action satisfies proper variation under the boundary conditions. The equations of motion for $q$ and $\vartheta$ are the same as mentioned in eq. (\ref{eq:eom_q_theta}). The set of coupled equations of motion can be solved for the chosen boundary conditions, yielding
\begin{equation}
\label{eq:background_axion-neumann}
    \begin{split}
        &q_\pm(t)=
\sqrt{
r^2(1-t)^2
+q_{1}^{\,2}t^2
\pm 2r_\pm q_{1}
\cosh(\Delta\vartheta)\,t(1-t)
},\\
& \vartheta_\pm(t)
=\frac12
\log\!\left(
\frac{
r_\pm(1-t)
\pm q_{1}e^{\Delta\vartheta}t
}{
r_\pm(1-t)
\pm q_{1}e^{-\Delta\vartheta}t
}
\right),\hspace{10mm} r_\pm=\pm q_{1}\cosh(\Delta\vartheta)-\alpha N_{c}
 \end{split}
\end{equation}
As before, we will stick to the physical $+$ branch and will ensure that $q(t)$ can never pass through zero to avoid $\vartheta(t)$ from any logarithmic singularity. Utilizing the above expressions for $q(t)$ and $\vartheta(t)$, one can compute the on-shell action. The axion path integral for fixed boundary value of axion $\Delta\vartheta$, evaluates to
\begin{equation}
\label{eq:fixed_charge_integral}
Z[\alpha,q_1;0,\Delta\vartheta]
\simeq
\int_{0}^{\infty} dN_c\,
\exp\!\left\{
i\,
\frac{3\pi^2}{2N_c\hbar}
\left[
N_c^2\left(4+\alpha^2\right)
-2N_c\alpha q_1\cosh(\Delta\vartheta)
+q_1^2\{\sinh\!\left(\Delta\vartheta\right)\}^2
\right]
\right\}.
\end{equation}
To evaluate the integral, we use the Bessel identity, 
$\int_0^\infty dx \exp(-a x-b/x)=2\sqrt{b}K_1(2\sqrt{a}\sqrt{b})/\sqrt{a}$, where ${\rm Re}(a)>0$ and ${\rm Re}(b)>0$. Considering the case when $(4+\alpha^2)>0$, we have

\begin{equation}
\begin{aligned}
Z[\alpha,q_1;0,\Delta\vartheta]
&\simeq
\,\frac{2q_1\sinh(\Delta\vartheta)}
{\sqrt{4+\alpha^2}}\,
e^{-3\pi^2 i \alpha q_1 \cosh(\Delta\vartheta)/\hbar}
\\
&\qquad\times
K_1\!\left(\epsilon{-}\,
3\pi^2 i\,q_1\sqrt{(4+\alpha^2)\{\sinh(\Delta\vartheta)\}^2}/\hbar\,
\right), \quad \ep>0
\end{aligned}
\end{equation} 
To reach the other side of the duality, one needs to take the Fourier transform of the above fixed $\Delta\vartheta$ path integral via $e^{in\Delta\theta}$. Taking the limit $\hbar\to 0$ and utilizing approximation $K_{1}(z)\sim \sqrt{\frac{\pi}{2}}\,e^{-z}/\sqrt{z}$,
we obtain (ignoring the Jacobian)

\begin{equation}
\begin{aligned}
&Z[\alpha,q_1;n]
\simeq
\int d\Delta\vartheta\,A e^{i\Phi_N/\hbar}
%\simeq
%\int d\Delta\vartheta\,
%e^{i\sqrt{3}\ n\,\Delta\vartheta/\sqrt{2}\hbar}
 %\exp\!\left[3\pi^2  q_1i
%\left(
%-\alpha\cosh(\Delta\vartheta)
%+\sqrt{(4+\alpha^2)\{\sinh(\Delta\vartheta)\}^2}
%\right)/\hbar
%\right] 
\\
\text{where,}\,\,\,&\Phi_N
=
6\pi^2
\left(
 \tilde{n}\,\Delta\vartheta
-\frac{1}{2}
\left[
\alpha q_1\cosh(\Delta\vartheta)
-
q_1\sqrt{(4+\alpha^2)\{\sinh(\Delta\vartheta)\}^2+i\ep}
\right]
\right)
\end{aligned}
\end{equation}
The presence of $i\epsilon$ prescription to determine the square root properly so that the branch points move slightly away from the real axis and $\lim\limits_{\ep\rightarrow 0}\sqrt{z^2+i\ep}=|z|$.
The saddle points and the action at the saddles are given by

\begin{equation}
\label{eq:action_neuman_Scalar}
\begin{split}
&\Delta\vartheta_{\sigma,\zeta}^{(k)}
=
\ln\!\biggl\{
-\sigma\,
\left(\sqrt{4+\alpha^{2}}+\sigma\alpha\right)
/
2
\biggr\}+\zeta\cosh^{-1}\biggl(\frac{\tilde{n}}{q_1}\biggr)
+2\pi i k,
\\
&
\Phi_N(\Delta\vartheta_{\sigma,\zeta}^{(k)})
=
6\pi^2
\biggl[
\tilde{n} \Delta\vartheta_{\sigma,\zeta}^{(k)}
-\zeta\sqrt{\tilde{n}^2-q_1^2}
\biggr],\qquad \sigma,\zeta\in\{+1,-1\},\,\,
k\in\mathbb{Z}.
\end{split}
\end{equation}
Although there are four saddles, not all of them will be in the principal sheet. Substituting the saddles in the on-shell solution for $\theta(t)$ as given in eq. (\ref{eq:background_axion-neumann}), we obtain the same scalar profile as found for Dirichlet in eq. (\ref{eq:scalar_axion_profile}). Comparing the above action in eq. (\ref{eq:action_neuman_Scalar}) to the action as found in the three-form case, mentioned in eq. (\ref{eq:action_at_saddle_form_Side}), we find that the duality holds if the relevant saddles on both sides are such that $(\rho,\eta;\sigma,\zeta;k)$ is either $(+1,+1;\,-1,-1;\,0)$ and/or $(+1,-1;\,-1,+1;\,0)$. We have also observed another possibility when no saddle is relevant on either side. In this case, the duality is trivially satisfied. 

\begin{figure}[htbp]
\centering
\subfigure[\,\,$0\leq|\alpha|<|\alpha|_c$ ]{
\includegraphics[width=0.48\textwidth]{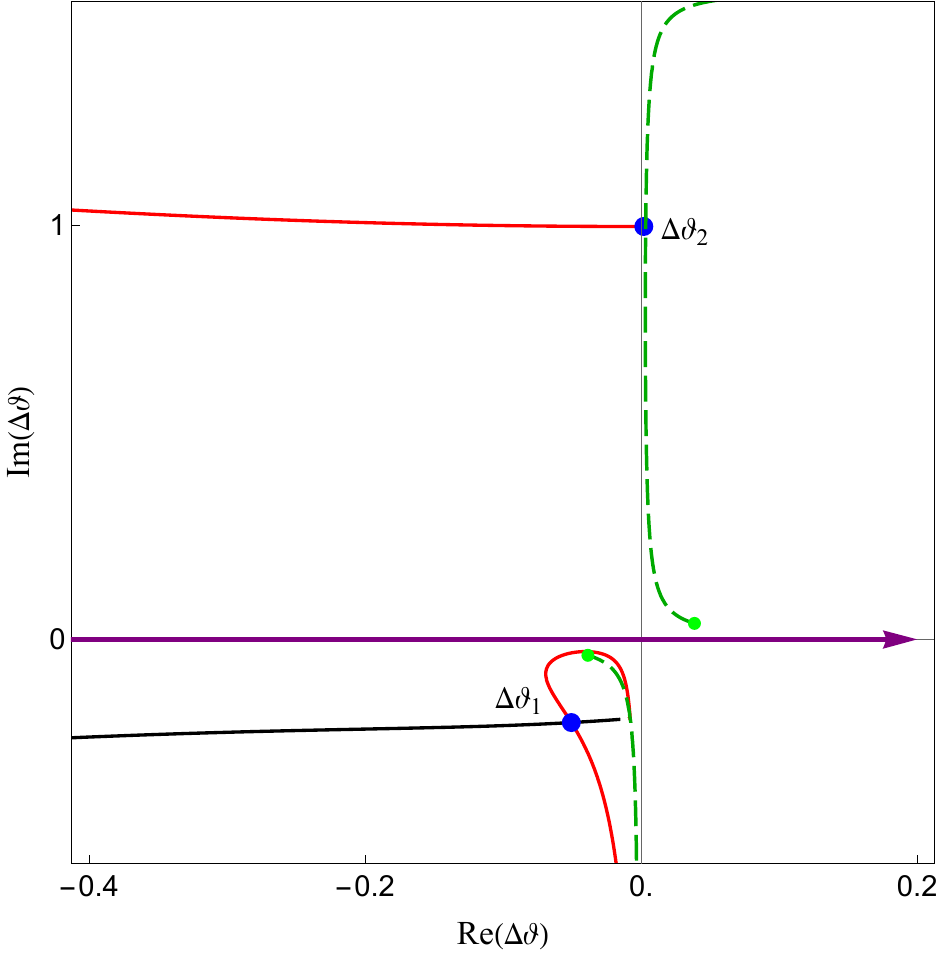}
}
\subfigure[\,\,$|\alpha|_c\leq|\alpha|<2$]{
\includegraphics[width=0.48\textwidth]{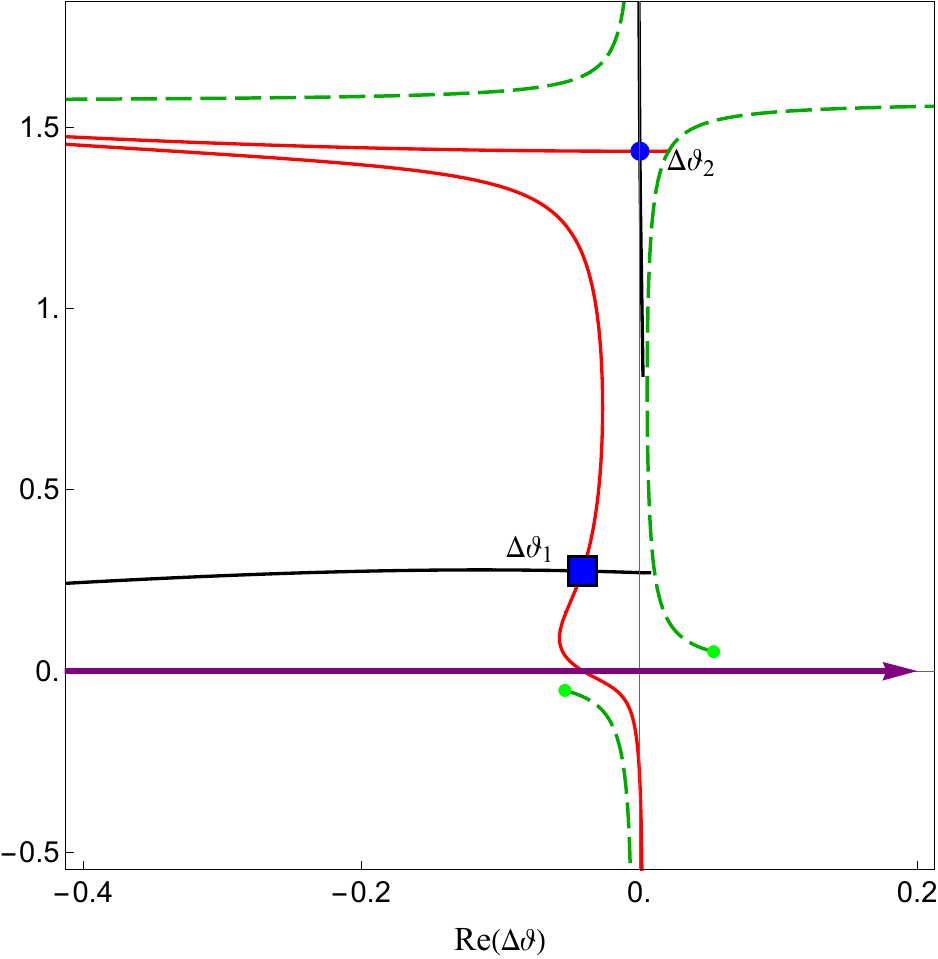}
}
\caption{PL plot for wormhole solutions in the range $0\leq |\alpha|<2$. For $0<|\alpha|<|\alpha|_c$ no saddle is relevant. For $|\alpha|_c<|\alpha|<2$, only one saddle is relevant ($\sigma=\zeta=-1$). We choose $ |\alpha|=0.8,$ $q_1=6,$ $\tilde{n}=5$, $ n=242$ in Left and $ |\alpha|=1.5,$ $q_1=6,$ $\tilde{n}=5,n=242$ in the right figure.}
\label{fig:axion-neumann}
\end{figure}

To analyze the duality, we perform the PL analysis in the cut-$\Delta\vartheta$ plane, which contains branch cuts, unlike the other side of the duality, where we did PL in the covering space using the physical lift of the contour. As shown in the PL plots for various regimes of $q_1$ and $\alpha$ in fig. \ref{fig:axion-neumann}, these conditions are satisfied by the relevant saddles on both sides of the duality. The cross-throat saddles always remain irrelevant while the same-side saddle becomes relevant for $|\alpha|>|\alpha|_c$, see table. \ref{tab:duality-saddle-choices}
\begin{table}[t]
    \centering
    \begin{tabular}{|c| c| c| c|}
        \hline
        Choice
        & Saddle type
        & $(\rho,\eta;\sigma,\zeta;k)$
        & PL relevance \\
        \hline
        I
        & Same-side
        & $(+1,+1;\,-1,-1;\,0)$
        & Relevant \\
        II
        & Cross-throat
        & $(+1,-1;\,-1,+1;\,0)$
        & Irrelevant \\
        \hline
    \end{tabular}
    \caption{The table summarizes the Picard-Lefschetz relevance of various
    same-side and cross-throat saddles across the range of parameters. This is for the choice of contour $N_c\in (0,\infty)$.}
    \label{tab:duality-saddle-choices}
\end{table}

Now, let us derive the bound on the imaginary value of the boundary axion ($\Delta\theta$). As we choose $\eta=-1$ on the other side of the duality, we then follow the table (\ref{tab:duality-saddle-choices}) and choose $\sigma=-1,\zeta=+1,k=0$. With this choices, in the asymptotic limit $q_1\rightarrow\infty$, we get from the saddles of $\Delta\vartheta$, ($\alpha=-i|\alpha|$, as we chose $\rho=+1$)
\begin{equation}
    \label{eq:asymp-theta-sad-neumann}
\Delta\vartheta_{-1,+1}^{(0)}\biggr|_{q_1\rightarrow\infty}=\ln\biggl(\frac{i\sqrt{4-|\alpha|^2}-|\alpha|}{2}\biggr)=i\biggl[\pi-\cos^{-1}\biggl(\frac{|\alpha|}{2}\biggr)\biggr]
\end{equation}
Restoring the normalization, the above equation is precisely the imaginary bound on $\Delta\theta$. For $|\alpha|=2$, we get
\begin{equation}
    \label{eq:IDB}
    |{\rm Im}(\Delta\theta_{-1,+1}^{(0)})|\biggr|_{|\alpha|=2}=2\tau_{\rm IDB}. 
\end{equation}
This is the same as obtained in eq. (\ref{eq:bound_on_theta_alpha}) by analyzing the convergence of the sum. This derives the bound directly from the saddles of the dual scalar axion. Hence, we conclude that the saddle corresponding to the cross-throat geometry is irrelevant when computing the path integral for the choice of contour $N_c\in(0,\infty)$.
\begin{figure}[htbp]
\centering
\subfigure[\,\,3-form]{
\includegraphics[width=0.47\textwidth]{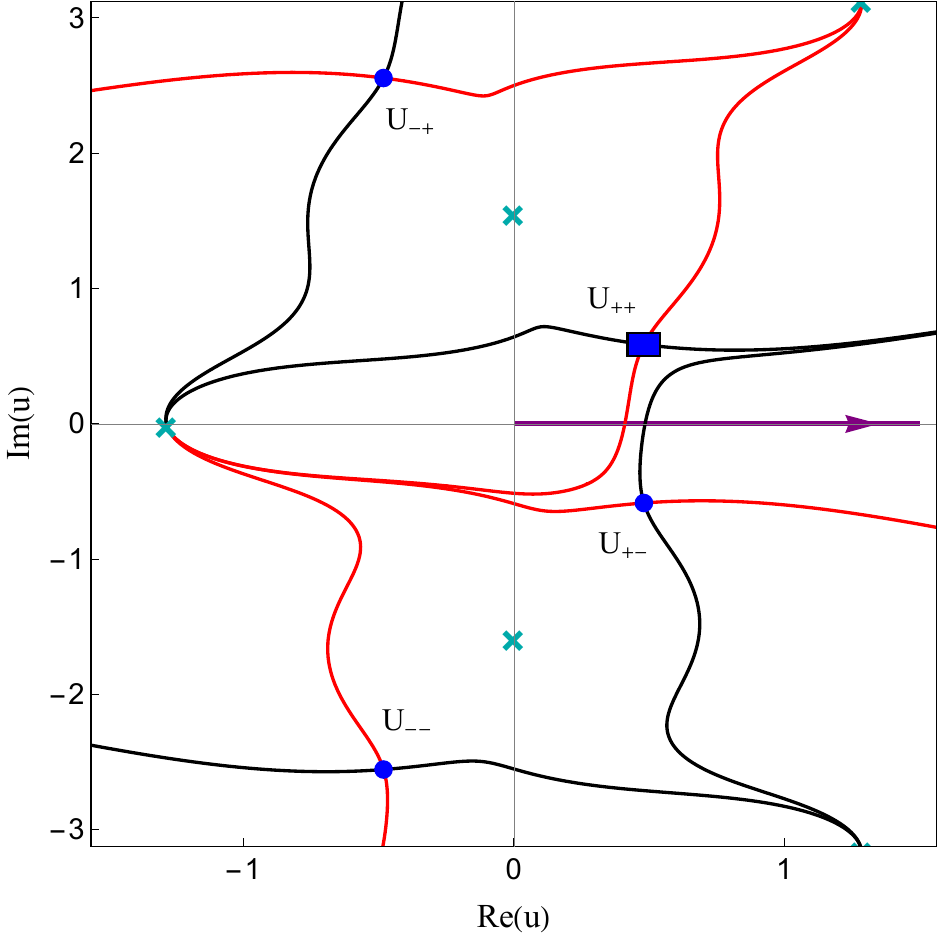}}
\subfigure[\,\,scalar]{
\includegraphics[width=0.47\textwidth]{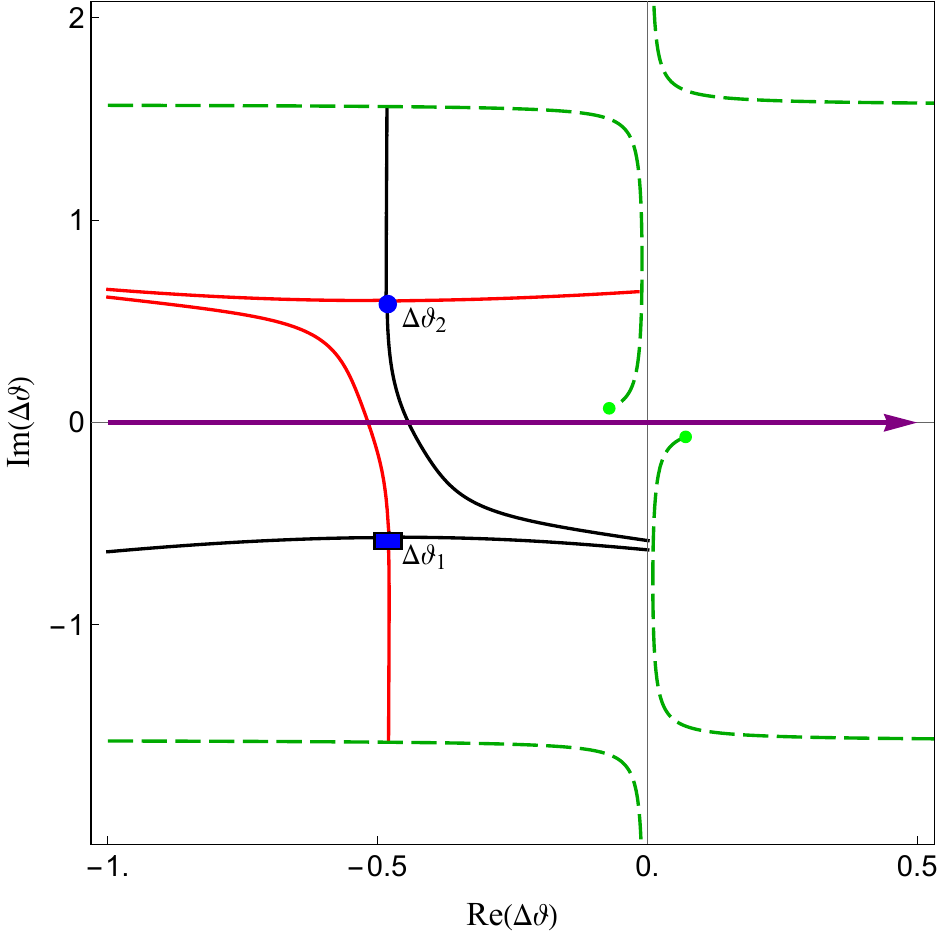}
}
\caption{PL plot for wormhole solutions for real values of $\alpha$ where $\alpha=1,$  $\tilde{n}=n/2\pi^2\sqrt{6},$ $ n=240,$  $ q_1=6$. Both in the three-form side and the scalar axion side of the duality. It explicitly demonstrates that the relevant saddle satisfies $(\rho,\eta,\sigma,\zeta)=$ $(+1,+1,-1,-1)$ as mentioned in table \ref{tab:duality-saddle-choices}}
\label{fig:real_alpha_plot}
\end{figure}

\begin{figure}[htbp]
\centering
\subfigure[\,\,3-form]{
\includegraphics[width=0.46\textwidth]{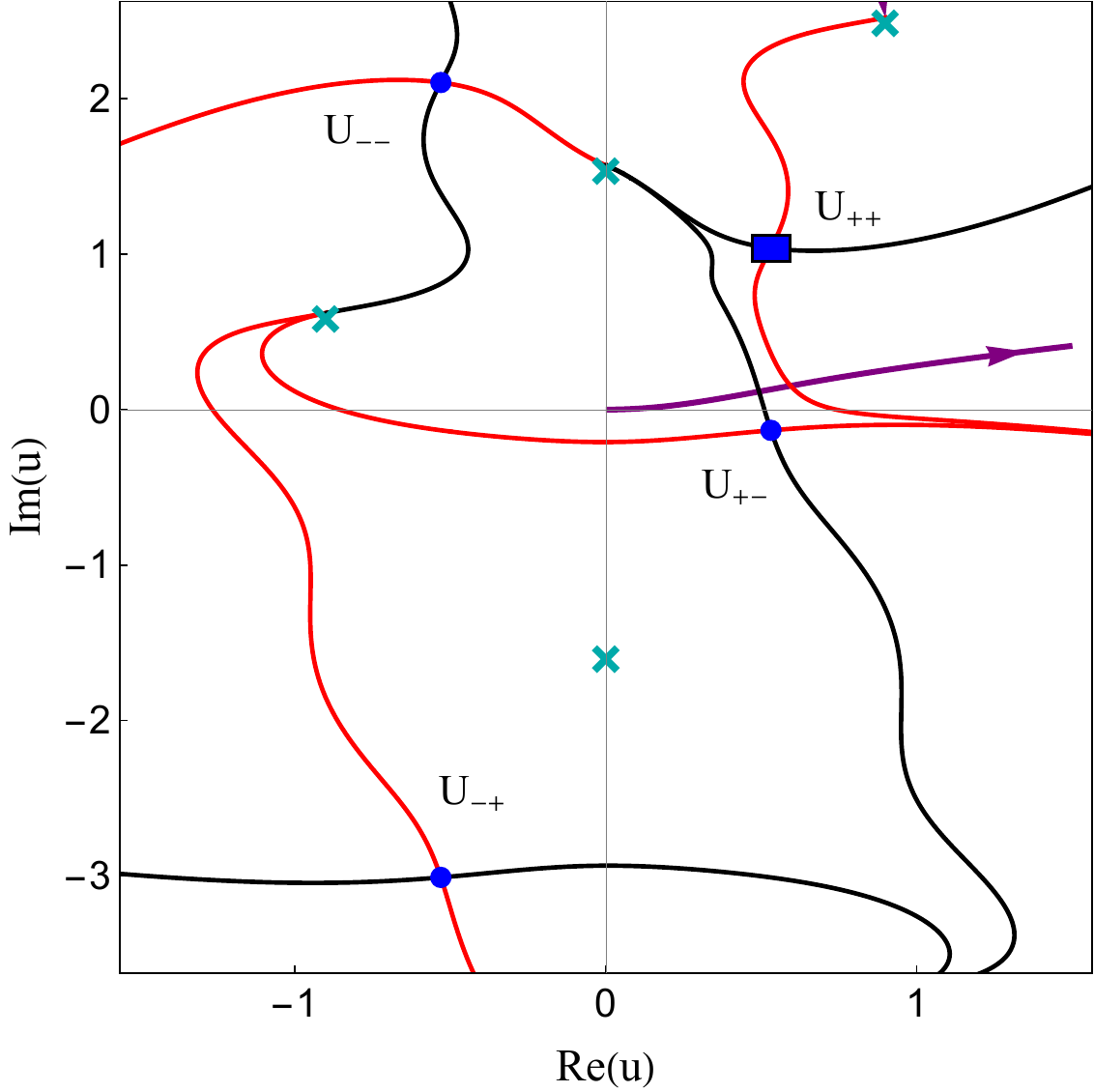}}
\subfigure[\,\,scalar]{
\includegraphics[width=0.46\textwidth]{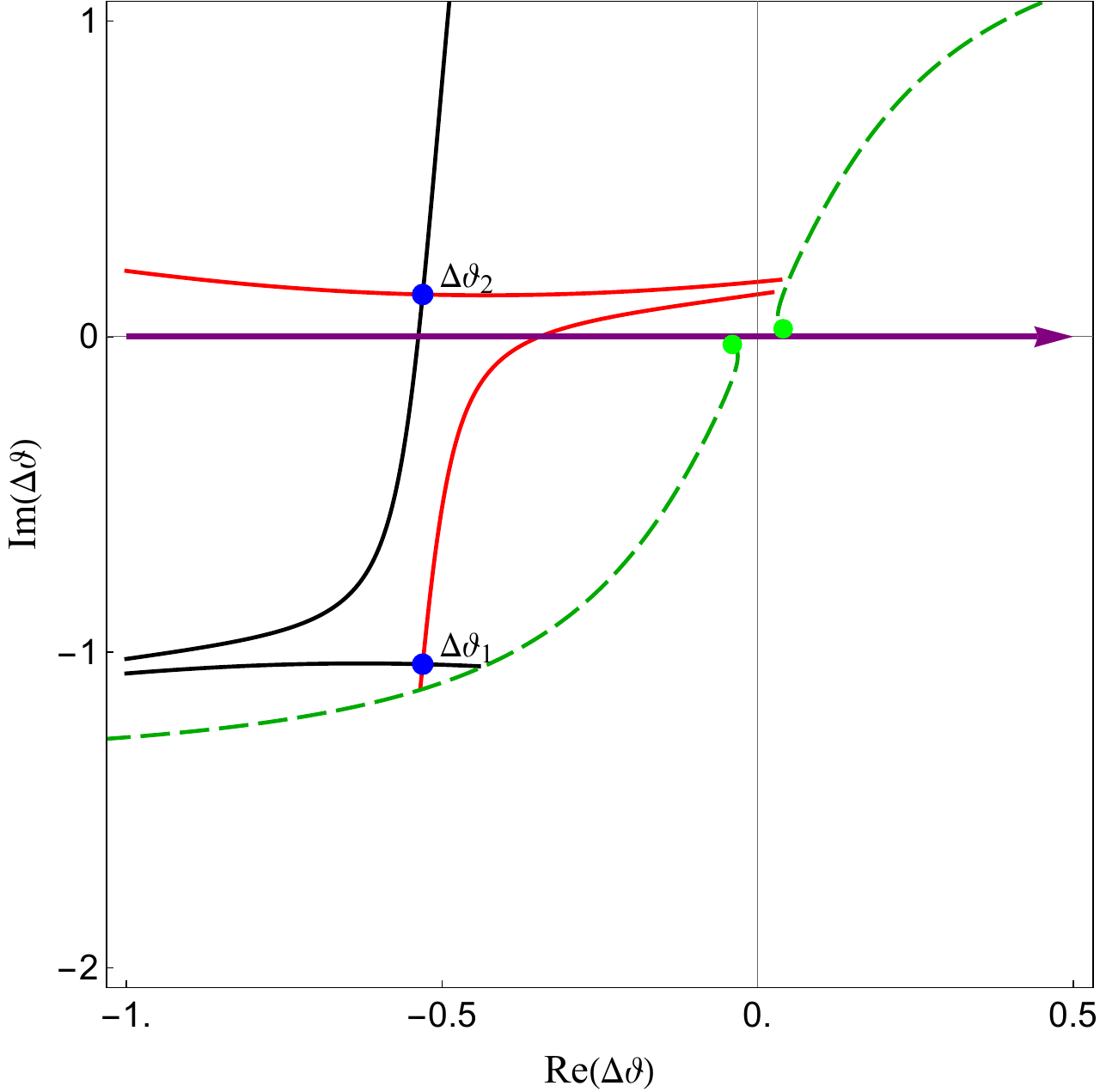}
}
\caption{PL plot for wormhole solutions for complex values of $\alpha$ where $\alpha=\sqrt{2}\exp(i\pi/4),$ $q_1=6,$ $\tilde{n}=n/2\pi^2\sqrt{6},$ $n=240$. Both in the three-form side and the scalar axion side of the duality. It explicitly demonstrates that the relevant saddle satisfies $(\rho,\eta,\sigma,\zeta)=$ $(+1,+1,-1,-1)$ as mentioned in table \ref{tab:duality-saddle-choices}}
\label{fig:complex_alpha_plot}
\end{figure}

%%%%%%%%%%%%%%%%%%%%%%%%%%%%%%%%%%%%%%%%%%%%%%%%%%%%%%%

\subsection{Complex saddles and duality transformation}
\label{eq:duality check}

In the preceding sections, we focused on Euclidean/imaginary wormholes and examined the scalar--two-form duality in these backgrounds. Such configurations play a central role in deriving the distance bound. As demonstrated in Sec. \ref{sec:duality_and_condition}, however, the duality is independent of the choice of background geometry: the stress tensors and equations of motion in the two formulations coincide for arbitrary values of the lapse $N_c$, including generic complex backgrounds, as summarized in appendix.(\ref{sec:stres_tensor_equality}). In this section, we provide a direct verification of this statement at the level of Lefschetz thimbles. As pointed out in table \ref{tab:dirichlet-neuman-saddle-relevance}, to obtain a complex saddle, we choose $\alpha$ to be real and generic complex. 

The Picard-Lefschetz plot for both real and complex $\alpha$ is illustrated for both sides of the duality in figs. \ref{fig:real_alpha_plot} and \ref{fig:complex_alpha_plot}, respectively. The relevant saddles are in perfect agreement with the duality and satisfy the table \ref{tab:duality-saddle-choices}. For complex saddles in the $u$-plane, the dual scalar is also complex, as required by duality.

%%%%%%%%%%%%%%%%%%%%%%%%%%%%%%%%%%%%%%%%%%%%%%%%

\section{The bound on $\alpha$, IDB and KSW criterion}
\label{sec:ksw}
Complex saddles play an important role in gravitational path integrals, but they have to be included with care. Otherwise, they may rise to unphysical configurations and pathologies in the path integral \cite{Jonas:2022uqb,Lehners:2021mah, Lehners:2022xds, Witten:2021nzp, Jonas:2022uqb, BenettiGenolini:2026raa, Guo:2026vkl}. It is therefore essential to assess their allowability using the Kontsevich-Segal-Witten (KSW) criterion \cite{Kontsevich:2021dmb, Witten:2021nzp}.
While Euclidean wormholes are always KSW-allowed, there are other regimes in which complex saddles arise. One such example is when $\alpha$ is purely imaginary but lies outside this range $0<|\alpha|<2$, see table \ref{tab:dirichlet-neuman-saddle-relevance}. It is therefore natural to ask what the allowable regime of $\alpha$ is according to the KSW criterion. We address this in this section. Secondly, the criterion also plays an important role in the discussion of imaginary distance bound (IDB). As noted in \cite{Maldacena:2026jqd}, the IDB and KSW criteria coincide in some cases, and the IDB was proposed as a generalization of KSW to all moduli. In this section, we aim to realize it explicitly within the minisuperspace framework. We want to reproduce the IDB inequality in eq.   (\ref{eq:bound_on_theta_alpha}) interpolating the half and full wormhole, directly from the principles of the KSW criterion.

Consider $g_{\mu\nu}(x)$ be a complex metric on a $\mathfrak{D}\geq2$ dimensional spacetime ($\mathcal{M}$). We define the Euclidean path integral of a $p$-form gauge field $A$ with the field strength $F=dA$, which is a $q=p+1$ form. According to the KSW allowability criterion, $g_{\mu\nu}$ is allowable, if it satisfies the condition \cite{Kontsevich:2021dmb,Witten:2021nzp}

\bea
\label{KSW}
\begin{split}
\mathcal{I}_q[A]& =\frac{1}{2q!}\int_M d^Dx\sqrt{det\,g}\,\,g^{\mu_1 \nu_1}\cdots g^{\mu_q \nu_q} F_{\mu_1 \mu_2...\mu_q} F_{\nu_1 \nu_2...\nu_q} \, ,\\
g_{\mu\nu} & \, \text{is KSW allowable iff} \, \,\, {\rm Re}\left(\sqrt{det\,g}\,\, g^{\mu_1 \nu_1}\cdots g^{\mu_q \nu_q}F_{\mu_1 \mu_2...\mu_q} F_{\nu_1 \nu_2...\nu_q}\right) > 0 \, ,
\end{split}
\eea
for all $q \in \{0, ... , D\}$, where $\mathcal{I}_q[A]$ is the Euclidean action. This follows directly from the convergence of the path integral $\int\mathcal{D}A\, e^{-\mathcal{I}_q[A]}$. For the metrics having a diagonal form, i.e., $g_{\mu\nu}=\lambda_{\mu}(x)\delta_{\mu\nu}$, the criterion simplifies to 
\beq
\label{eq:KSW_criterion}
\Sigma(x)\equiv \sum_{\mu=0}^{D-1}|\arg \, \lam_{\mu}(x)| < \pi \quad \forall x \in \mathcal{M}\, ,
\eeq
where $\arg (z) \in (-\pi,\pi]$ with $z$ is complex. For the diagonal metric ansatz as in eq. (\ref{eq:metric}), the left-hand side of the inequality becomes

\begin{equation}
    \label{eq:ksw_sigma}
    \Sigma(t)=\biggl|\arg\biggl(\frac{-N_c^2}{q(t)}\biggr)\biggr|+3|\arg(q(t))|\equiv \Sigma_{\rm temporal}(t)+\Sigma_{\rm spatial}(t).
\end{equation}
For the minisuperspace background $\Sigma$ depends only on $t$. Observing that both the $\Sigma_{\rm temporal}$ and $\Sigma_{\rm spatial}$ are non-negative, a metric will be KSW disallowed if
\begin{equation}
    \label{eq:weak KSW}
    |\Sigma_{\rm spatial}(t)|\geq\pi\quad \text{for any $t\in[0,1]$}. 
\end{equation}
However, $|\Sigma_{\rm spatial}(t)|<\pi$ doesn't guarantee the KSW allowability. This is a weaker criterion that analyses only the spatial component of the KSW argument \cite{Ailiga:2025fny, Ailiga:2025osa}. We will analyze this below.

Consider the geometries at $t=0$. The initial size of the saddle geometries (eq. (\ref{eq:saddle_expression})) is given by 
\begin{equation}
    \label{eq:initial_size}
    q(0)=\rho \frac{2\tilde{n}}{\sqrt{4+\alpha^2}},\quad \rho=\pm 1.
\end{equation}
For $\alpha=i\epsilon|\alpha|$ with $|\alpha|>2$, the geometries become purely imaginary and hence $\Sigma_{\rm spatial}(0)=3\pi/2>\pi$. Hence, when we extend the absolute value of $\alpha$ beyond 2, the geometry becomes KSW-disallowed. Moreover, for $|\alpha|<2$, one requires $\rho=+1$, otherwise it will also violate the KSW requirement. It is interesting to note from table (\ref{tab:duality-saddle-choices}) that this is the same sign as also required for the duality to hold. This explicitly shows the agreement between duality and the KSW criterion. Below, we consider only $\rho=+1$.

\begin{figure}
\includegraphics[width=0.65\linewidth]{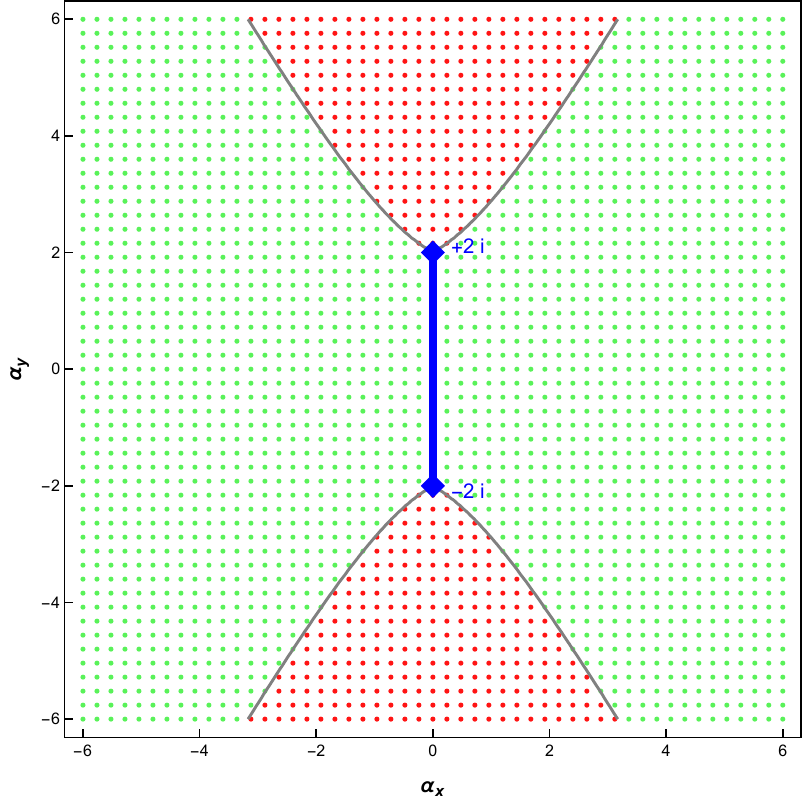}
    \caption{The plot shows the KSW allowability of the Neumann boundary condition ($\alpha$). The red region is disallowed by the KSW criterion. The green region is allowed only by the weak KSW criterion. The blue segment shows the real and Euclidean wormhole solutions: $\alpha=\pm i|\alpha|,|\alpha|\leq 2$. The points at $\pm 2i$ are the ``imaginary wormholes" solutions satisfying the IDB. These lie on the boundary of the permissible boundary condition. }
    \label{fig:KSW_allowed_alpha}
\end{figure}

\textbf{Generic complex $\alpha$.}
For arbitrary $\alpha\in\mathbb C$, we write $\alpha=\alpha_x+i\alpha_y$. From eq. (\ref{eq:initial_size}), we get the following condition for the geometry to be KSW disallowed
\begin{equation}
\label{eq:KSW_disallowedness}
|\arg(4+\alpha^2)|\geq\frac{2\pi}{3}.
\end{equation}
The above equation puts a strong constraint on the allowed Neumann boundary condition. The allowed/disallowed parameter regimes are shown in figure \ref{fig:KSW_allowed_alpha}. Note that in the blue segment ($\alpha=i|\alpha|,\,|\alpha|<2$), we have real and Euclidean wormholes. The endpoints of this segment describe the ``imaginary wormholes" satisfying the imaginary distance bound on the boundary value of the axion, as explained in sec. \ref{eq:neumann-3-form-side}. The boundary of the two regions ($\mathbb{D}_{\rm Bd}^{\rm KSW}$) is given by 
\begin{equation}
    \label{eq:allowable_boundary}
    \mathbb{D}_{\rm Bd}^{\rm KSW}=\{(\alpha_x,\alpha_y):\sqrt{3}|\alpha_y| =| \alpha_x| +2 \sqrt{\alpha_x^2+3}\}.
\end{equation}
Also note that the allowed-disallowed region is symmetric around the real and imaginary axes ($\alpha_x\rightarrow-\alpha_x,\alpha_y\rightarrow-\alpha_y$). Similar symmetry has been observed earlier in \cite{Ailiga:2025fny}.

\begin{figure}
    \subfigure[\,IDB]{\includegraphics[width=0.45\linewidth]{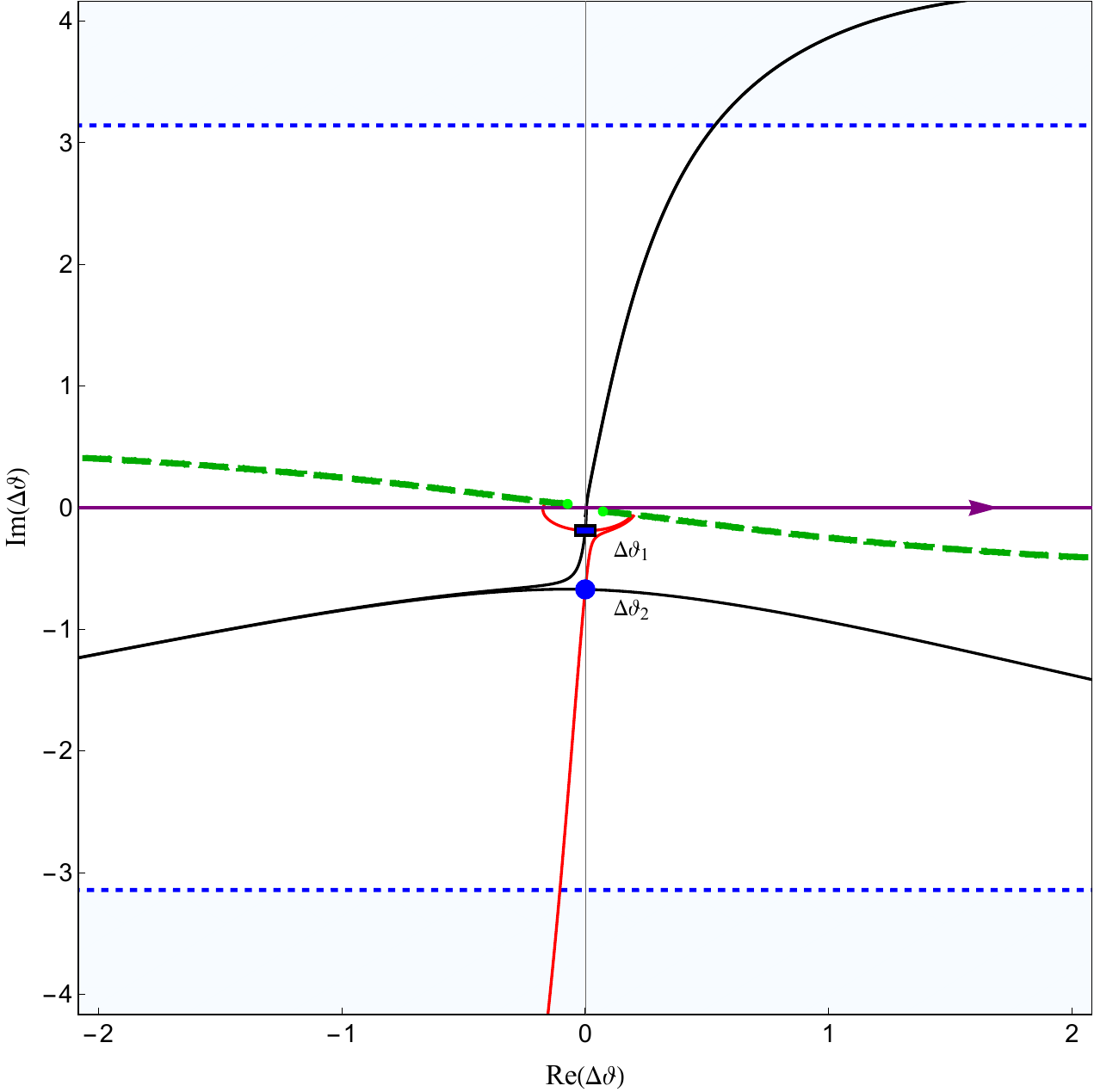}}
    \subfigure[\,KSW]{\includegraphics[width=0.45\linewidth]{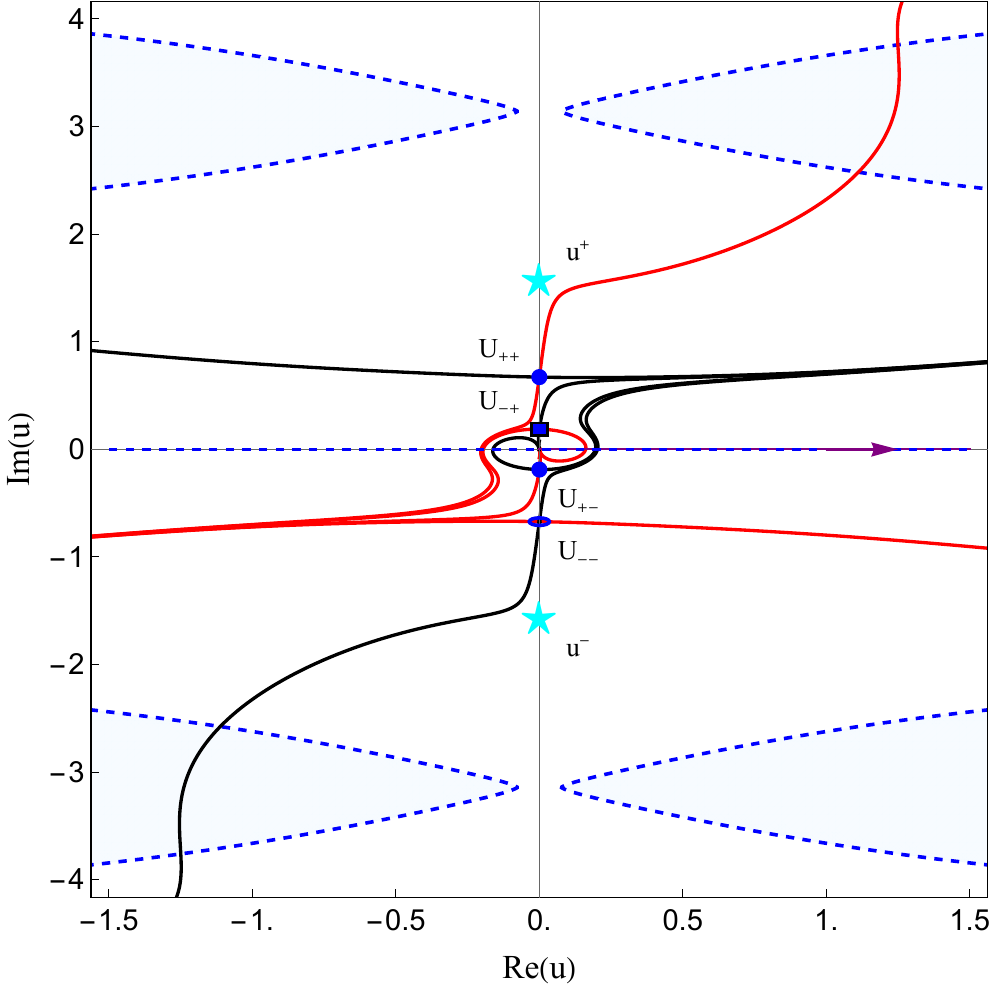}}
    \caption{The plot showing the cropping of the Lefschetz thimbles (the part which is inside the blue shaded region) (a) in the complex $\Delta\vartheta$ plane by the imaginary distance bound ($|{\rm Im}(\Delta\vartheta)|<\pi$) (b) in the covering $u$-plane for fixed metric boundary condition. Similar cropping of thimbles due to the KSW has also been observed previously \cite{Lehners:2021mah, Jonas:2022uqb}. Here, we find that the similar cropping of thimbles also happens for the $\Delta\theta$ plane. This shows the compatibility between the IDB and KSW criterion.}
   \label{fig:IDB_cropping_thimble}
\end{figure}
Having found the bound on the allowed parameter space ($\alpha$) where Euclidean solutions exist, we now proceed to compute the inequality relating the imaginary part of $\Delta\theta$ and $\alpha$, as given in eq. (\ref{eq:bound_on_theta_alpha}) directly from the KSW criterion. To derive the inequality, use inputs from the duality. The stationary point of the integral as appeared in the scalar side of the duality in eq. (\ref{eq:fixed_charge_integral}) are $N_\lam^*=\lam q_1\sinh(\Delta\vartheta)/\sqrt{4+\alpha^2},\lam=\pm 1$. The initial value $\mathbf{r}$, as mentioned in eq. (\ref{eq:background_axion-neumann}), at these saddles are given by ($\alpha=i\ep|\alpha|,\ep=\pm 1$)

\begin{equation}
    \label{eq:bound from KSW_r}
   \mathbf{r}_{\lam\ep}^* = q_{1}\cosh(\Delta\vartheta)-\alpha N_{\lam}^*=q_1\cosh(\Delta\vartheta)-i\ep\lam\frac{q_1|\alpha|\sinh(\Delta\vartheta)}{\sqrt{4-|\alpha|^2}}.
\end{equation}
Now we use the following key input from the duality: although $\Delta\vartheta$ can be complex in the regime where the saddles are complex, it is purely imaginary in the regime where we have Euclidean wormholes as required by the duality, see sec. (\ref{sec:stres_tensor_equality}) in appendix. On writing $\Delta\vartheta=iy$, we get from eq. (\ref{eq:bound from KSW_r})
\begin{equation}
    \label{eq:bound_on_alpha_ksw}
    \mathbf{r}_{\lam\ep}^* = q_{1}\cos( y)+\ep\lam\frac{q_1|\alpha|\sin(y)}{\sqrt{4-|\alpha|^2}}=\frac{q_1}{\sin(\beta)}\sin(\boldsymbol{\beta}+\ep\lam y),
\end{equation}
where $\boldsymbol{\beta}=\cos^{-1}(|\alpha|/2)$. As already explained, we must have $\mathbf{r}_{\lam\ep}^*>0$. This is equivalent to
\begin{equation}
    \label{eq:sign_of_r_ksw}
    \begin{split}
        &\ep\lam=+1 \quad \Rightarrow \quad -\boldsymbol{\beta}<y<\pi-\boldsymbol{\beta},\\
        &\ep\lam=-1 \quad \Rightarrow \quad \boldsymbol{\beta}-\pi<y<\boldsymbol{\beta}.
    \end{split}
\end{equation}
Therefore, there exists at least one KSW allowed scalar-frame saddle, if and only if one satisfies
\begin{equation}
    \label{eq:cond_scalar_saddle}
    |y|<\pi-\boldsymbol{\beta}.
\end{equation}
Substituting $\boldsymbol{\beta}$ and $\Delta\vartheta=\sqrt{2/3}\Delta\theta$ in eq. (\ref{eq:cond_scalar_saddle}), We arrive at the following inequality
\begin{equation}
    \label{eq:bound from KSW}
    \left|\operatorname{Im}\Delta\theta\right|+\sqrt{\frac{3}{2}}\arccos\left(\frac{|\alpha|}{2}\right)<
\sqrt{\frac{3}{2}}\pi,
\end{equation}
which is precisely eq. (\ref{eq:bound_on_theta_alpha}). Hence, we conclude that the bound on $\Delta\theta$ and $|\alpha|$ is also consistent with KSW and can be reproduced independently. This further highlights the compatibility between the imaginary distance bound and the KSW criterion across different metric boundary conditions. It is worth emphasizing that, although we have imposed only a restricted, or ``weak,'' KSW condition evaluated at $(t=0)$, this condition is already sufficiently restrictive to reproduce the complete bound. This is because $\mathbf{r}>0$ forces us to stick to the physical sheet.
To get the imaginary distance bound, we set $|\alpha|=2$ in the above inequality and obtain $|{\rm Im}\Delta\theta|<\sqrt{3/2}\pi$, which is precisely $2\tau_{\rm IDB}$ as defined in \cite{Maldacena:2026jqd, DiUbaldo:2026rly}.

Let us now examine how the imaginary distance bound constrains the Lefschetz thimbles in the $\Delta\vartheta$ plane. As illustrated in the fig. \ref{fig:IDB_cropping_thimble} the steepest descent flowlines from the relevant saddle get cropped by the horizontal lines at $|{\rm Im}(\Delta\vartheta)|=\pi$ imposed by the distance bound in eq. (\ref{eq:IDB}). Once the distance bound is imposed, only the portions of these contours lying within the allowed strip remain admissible; the integration cycles are therefore effectively truncated at the boundaries. This closely resembles the truncation of Lefschetz thimbles upon entering the domain of KSW non-allowable metrics \cite{Lehners:2021mah, Jonas:2022uqb}. We suspect this is a different manifestation of the same underlying mechanism. 

To obtain the KSW disallowed region, in the three-form side of the duality, we consider the $q_+(t)$ branch in eq. (\ref{eq:q(t)_dirichlet}). Substituting $N_c=q_0q_1\sinh(u)/2\tilde{n}$, we obtain $q_+(t;u)=\{q_0^2(1-t)^2+q_1^2t^2+2q_0q_1\cosh(u) t(1-t)\}^{1/2}$. Although $t\in[0,1]$, one can consider a deformed time-path in the complex $t$-plane where the endpoints are fixed. Each of these time paths will give rise to different $\Sigma(t)$. A geometry will be KSW allowable if there exists at least one such time path for which $\Sigma(t)<\pi$. To figure out the existence of any such time-path, we note the following topological argument: any such continuous path one can think of will necessarily cut the vertical axis $t=1/2+i\gamma$, where $\gamma\in\mathbb{R}$ \cite{Jonas:2022uqb}. Hence, it suffices to analyse the allowability on this infinite line. Then, by analyzing the spatial part of the KSW argument, one arrives at the following criterion for a geometry in the $u$-plane to be KSW non-allowable
\begin{equation}
    \label{eq:ksw_non-allowable_argument}
   \min\limits_{\gamma\in \mathbb{R}}\, 3\biggl|\arg q\biggl (\frac{1}{2}+i\gamma;u\biggr)\biggr| \geq\pi.
\end{equation}

As shown in the appendix \ref{sec:KSW_allowability_append}, the above inequality yields the following inequality in the complex $u=u_x+iu_y$-plane
\begin{equation}
    \label{eq:KSW_allowability_u_plane}
    \sqrt{3}\cosh(u_x)\cos(u_y)+|\sinh(u_x)\sin(u_y)|\leq -\sqrt{4\biggl(\frac{q_0^2+q_1^2}{2q_0q_1}\biggr)^2-1}.
\end{equation}
If the above inequality is satisfied, then there is no complex time path connecting $t=0$ and $t=1$ along which one can make the geometry KSW-allowable. Finally, we note that the saddles in the $N_c$-plane are KSW allowed while respecting duality in the $\Delta\theta$-plane, and the saddles satisfy the IDB. This happens for both the complex and Euclidean saddles.

%%%%%%%%%%%%%%%%%%%%%%%%%%%%%%%%%%%%%%

\section{Conclusion and Outlook}
\label{sec:conclusion}
%%%%%%%%%%%%%%%%%%%%%%%%%%%%%%%%%%%%%%%%%%%%%%%%%%

In this work, we investigate the duality between a massless axion $(\theta)$ and a two-form gauge field $(B)$ with three-form field strength $(H=dB)$ in four-dimensional spacetime \cite{Giddings:1987cg, Witten:2026twr}, within the Lorentzian minisuperspace path integral. We analyze the saddle-point geometries associated with Euclidean and/or complex axionic wormholes under both Dirichlet and Neumann boundary conditions. The Dirichlet problem for asymptotically flat wormholes was previously studied within the Picard-Lefschetz (PL) framework in \cite{Loges:2022nuw}. First, we extend their analysis by identifying the imaginary-wormhole saddles \cite{Maldacena:2026jqd, DiUbaldo:2026rly} and examining the configurations that determine the imaginary distance bound (IDB) in both the three-form and scalar sides of the duality. We have analyzed the relevance of imaginary wormhole saddles for the choice of Lorentzian contour using both the exact path integral and the PL in covering plane of lapse. We then extend the construction to a one-parameter family of Neumann boundary conditions that interpolates between a half-wormhole and the complete two-sided wormhole on both sides of the duality. Imposing the convergence of the sum over fixed charged wormholes, we derive an interpolating bound on the imaginary part of the boundary axion. An independent computation based on the Kontsevich-Segal-Witten (KSW) allowability criterion reproduces precisely the same bound, showing its compatibility with the IDB. We also analysed the duality across different complex saddles, demonstrating its background independence. We also found that the relevant saddles on both sides of the duality are admissible under the KSW criterion.

We first analyze the axion-three-form duality at the level of the minisuperspace path integral, keeping the boundary conditions on the metric generic. Once the axion is assigned a Dirichlet condition, and the dual three-form is treated with a Neumann condition, the scalar path integral is Gaussian. The homogeneous-mode determinant cancels the corresponding factor generated by the Fourier transform over the axion separation, and the two descriptions agree for reciprocal couplings, \(f_{\theta}=f_h^{-1}\) \cite{Witten:2026twr}. Under the duality transformation, the stress-tensor and the equation of motion also become identical. This establishes that the duality is independent of the metric boundary condition and holds on any generic (complex) background. Since the duality depends on the quantization condition, the three-form flux is integer quantized ($n\in\mathbb{Z}$) while the axion is periodic with period $2\pi$. This allows us to consider the sum over ($n$), which in the dual description relates to the sum over winding over $S^1$. In our analysis, the sum over flux sectors $n$ plays a central role, as the requirement of its convergence yields the imaginary distance bound.

We then impose Dirichlet boundary conditions on the metric and evaluate the path integral exactly. Its semiclassical expansion contains several distinct exponential contributions, which we identify with the relevant saddle geometries through a Picard–Lefschetz analysis. In the Euclidean regime, the saddles describe two different types of geometries- same-side and cross-throat. These are different segments of the full Giddings-Strominger (GS) wormhole. The same-side geometry is monotonic, while the cross-throat geometry passes through the throat. An analysis using the PL method in the covering coordinate reveals that the cross-throat saddles are always irrelevant to the path integral for the choice of Lorentzian contour ($N_c\in(0,\infty)$). Then we analyse the asymptotic behaviour of the saddle geometries. After subtracting the divergent arising from the asymptotic flat background, we discover that the action at the cross-throat saddle $(N_{--})$ becomes identical to the imaginary wormholes as considered in \cite{Maldacena:2026jqd, DiUbaldo:2026rly}. We also confirm that the saddle is absent in the semiclassical analysis of the exact path integral. Then, by analyzing the dual formulation in terms of the axion scalar, we found that at the dual saddle point, it satisfied the imaginary distance bound. However, the saddle is irrelevant according to the PL. This agrees with the analysis based on the convergence of the flux sum. This shows that the imaginary wormhole saddle corresponding to the distance bound is not relevant in the Lorentzian path integral.

Then we extend the construction to a one-parameter ($\alpha$) family of Neumann conditions at $t=0$, keeping the boundary condition to be Dirichlet at $t=1$. We observe that the Neumann boundary condition remains invariant under the metric rescaling, unlike the Dirichlet condition, which gives rise to distinctive features in the asymptotic regime. By restricting the parameter to be bounded in the regime $\alpha=i\ep|\alpha|,0<|\alpha|<2,\ep=\pm 1$, we realized the Euclidean wormhole geometries that describe various segments of the full GS wormhole, similar to the Dirichlet. The endpoints of this bound describe a half-wormhole ($|\alpha|=0$) and an imaginary wormhole ($|\alpha|=2$). A similar analysis utilizing the Picard-Lefschetz method in the covering coordinate, reveals that for values $|\alpha|<|\alpha|_c$, there is no relevant saddle while for $|\alpha|\geq |\alpha|_c$, one saddle becomes relevant, which describes the `` same-side" geometry. Hence, similar to the Dirichlet case, the cross-throat saddles are irrelevant and don't contribute to the path integral. Furthermore, we have analytically shown that these cross-throat saddles move to the unphysical sheet. Since the momentum condition is invariant under metric rescaling, no background subtraction of the divergences is required at $t=0$. Subtracting the divergences at $t=1$, we get a similar bound between the imaginary part of $\Delta\theta$ and $\alpha$. For $\alpha=2$, it yields the correct imaginary distance bound for the imaginary wormholes, while for $\alpha=0$ describing a half wormhole, we get half of the bound. We then reproduce the same bounds by analyzing the saddles of the scalar axion in the large boundary limit. The renormalized fixed-\(\Delta\theta\) on-shell action itself vanishes, in agreement with the scaling argument; the nonzero finite term in the fixed-flux frame arises through the Fourier/Legendre term.

After analyzing the Euclidean wormholes, we consider saddles that describe the complex background geometries. This happens when $\alpha$ is outside the above bounded region. As we showed, under the duality, both the stress tensor and the equation of motion are identical on both sides for a generic lapse ($N_c$). This prompts us to explicitly check the duality on such a generic complex background using the Picard-Lefschetz method. 
A direct comparison of actions and ascent cycles shows that the relevant saddle choices match between the two dual descriptions, even though their complete saddle sets need not. This demonstrates a nontrivial thimble-level check of the duality's background independence. 

Finally, we have analyzed the Kontsevich-Segal-Witten (KSW) allowability criterion on the complex Neumann parameter $\alpha$. By analyzing the spatial version of the criterion, we have independently found that the real Euclidean wormhole family lies on the segment \(\alpha=i|\alpha|\),\(|\alpha|\leq2\), and that is the only allowed configuration. Going beyond this range makes the saddle geometries disallowed. Furthermore, by considering the $\alpha$ to be a generic complex parameter, we constrain its allowability using the KSW criterion. We find that only certain $\alpha$ values define the complex geometries that are KSW-allowable. This imposes a strong constraint on the admissible space of boundary conditions. Then we derive the inequality in eq. (\ref{eq:bound_on_theta_alpha}) relating the imaginary part of the boundary axion to $|\alpha|$ using the KSW allowability principle. This highlights the compatibility between the imaginary distance bound and the KSW criterion, as noted in \cite{Maldacena:2026jqd}. Upon taking $|\alpha|=2$, we recover the imaginary distance bound ($\tau_{\rm IDB}$). In \cite{Lehners:2021mah, Jonas:2022uqb}, it was observed that thimbles get cropped because of the KSW criterion. Here, we found that, upon imposing the IDB bound, the thimbles are also cropped. Throughout our analysis, we find the duality consistent with the KSW criterion.

In this work, we have focused on asymptotically flat wormholes in the minisuperspace setup. A natural extension would be to consider asymptotically AdS axionic wormholes within the Lorentzian path-integral formalism. Another promising direction is to extend the analysis to axion-dilaton wormholes \cite{Andriolo:2022rxc, Jonas:2023ipa, Jonas:2023qle,Hertog:2024nys} and determine their contributing saddles through a systematic Picard-Lefschetz analysis. It will also be worthwhile to look for the contour of integration from a more fundamental principle. Finally, a comprehensive study of KSW allowability across the full complex saddle geometries and its relation to the imaginary distance bound remains an important avenue for future work.

\section{Acknowledgment}
We gratefully acknowledge the organizers of the workshop ``Advances in Black Hole Theory,'' held at IISER Pune, where some of the preliminary discussions that contributed to this work took place. AI has been used for language-editing and literature survey which were independently verified.

\section{Appendix}
\label{sec:apendix}
%%%%%%%%%%%%%%%%%%%%%%%%%%%%%%%%%%%%%%%%%%%

\subsection{Equality of the Stress Tensors under duality}
\label{sec:stres_tensor_equality}
In this appendix, we compute the stress-energy tensors of the scalar axion and the three-form field and show that they coincide upon imposing the duality relation. Consequently, the corresponding equations of motion are mapped into one another under the same duality transformation on a generic background/lapse ($N_c$).

Consider the following Lorentzian actions for a massless axion and the three-form field (we will set the coupling to unity $f_\theta=f_h=1$.)
\begin{equation}
    \label{eq:action_appendi}
    S_\theta=-\frac{1}{2}\int d^4x \sqrt{-g} g^{\mu\nu}\partial_\mu\theta\partial_\nu\theta,\quad S_H=-\frac{1}{2\, 3!}\int d^4x \sqrt{-g} \,H_{\alpha\beta\gamma}H^{\alpha\beta\gamma}.
\end{equation}
On defining the stress-tensors via $T_{\mu\nu}=-\frac{2}{\sqrt{-g}}\frac{\delta S}{\delta g^{\mu\nu}}$, we obtain the following stress tensors for the scalar and 3-form field, respectively
\begin{equation}
    \label{eq:stress-tensor}
    T_{\mu\nu}^{(\theta)}=\partial_\mu\theta\partial_\nu\theta-\frac{1}{2}g_{\mu\nu}(\partial\theta)^2,\quad T_{\mu\nu}^{(H)}=\frac{1}{2}H_{\mu\alpha\beta}H_{\nu}^{\,\,\alpha\beta}-\frac{1}{12}g_{\mu\nu}H^2.
\end{equation}
For the minisuperspace ansatz in eq. (\ref{eq:metric}), we have $g_{tt}=-N_c^2/q$, $g_{ij}=q\gamma_{ij}$. Together with an ansatz for the three-form field $H_{ijk}=h\,\epsilon_{ijk}, H_{tij}=0$, $\ep_{ijk}$ is Levi-civita tensor, we get the following components of the stress tensor
\begin{equation}
    \label{eq:component_stress-tensor}
    \begin{split}
        &T_{tt}^{(\theta)}=\frac{\dot{\theta}^2}{2},\quad T_{ij}^{(\theta)}=\frac{q^2\dot{\theta}^2}{2N_c^2}\gamma_{ij},\quad T_{ti}^{(\theta)}=0,\\
        &T_{tt}^{(H)}=\frac{h^2N_c^2}{2 q^4},\quad T_{ij}^{(H)}=\frac{h^2}{2q^2}\gamma_{ij},\quad T_{ti}^{(H)}=0,
    \end{split}
\end{equation}
where $\gamma_{ij}$ is the metric on the unit 3-sphere. Note that to get the same stress tensor on both sides, we require
\begin{equation}
    \label{eq:stress-tensor_equality}
T_{\mu\nu}^{(\theta)}=T_{\mu\nu}^{(H)}\qquad\Rightarrow\qquad \frac{\dot{\theta}}{N_c}=\frac{h}{q^2},\qquad \forall N_c.
\end{equation}
Note that the stress tensors coincide for arbitrary $(N_c)$, demonstrating that the duality holds independently of the choice of background geometry. In the language of differential form, this is $d\theta=*H$, where $*$ is the Hodge star operator \cite{Witten:2026twr}. It is easy to restore the couplings in the above duality transformation by simple rescaling $\theta\rightarrow\theta/f_\theta$ and $H_{\alpha\beta\gamma}\rightarrow H_{\alpha\beta\gamma}/f_h$. The condition is precisely the same as $\frac{d}{dt}(\dot{\theta} q^2)=0$ in eq. \ref{eq:eom_q_theta} upto normalization. The normalization can be fixed by demanding that $p_\theta=n$, where $p_\theta=2\pi^2\times q^2\dot{\theta}/N_c$, where $n=2\pi^2 h$. This agrees with the normalization of the multiplicative factor $e^{in\theta}$ as well. Under the same transformation in eq. (\ref{eq:stress-tensor_equality}), the equations of motion on both sides also become the same (eq. (\ref{eq:on-shell}) and \ref{eq:eom_q_theta}):

\begin{equation}
    \label{eq:quation_of_motion}
    \ddot{q}+\frac{2}{3}q\dot{\theta}^2=0 \qquad \leftrightarrow \qquad \ddot{q}+\frac{4\tilde{n}^2N_c^2}{q^3}=0,\qquad \forall N_c
\end{equation}
where, we defined $h=\sqrt{6}\,\tilde{n}$. It is worth commenting that the duality transformation in eq. (\ref{eq:stress-tensor_equality}) implies for real three-form $(h)$, $\theta$ has to be imaginary as $N_c$ is imaginary for the wormhole solutions. For $\theta=i\chi,\,\chi\in\mathbb{R}$ and $N_c=iN_E,\, N_E\in\mathbb{R}$, we get from the duality transformation eq. \ref{eq:stress-tensor_equality}
\begin{equation}
    \label{eq:complex scalar}
  \text{Imaginary wormhole:}\,\, \Rightarrow\quad  \frac{\dot{\chi}}{N_E}=\frac{h}{q^2}.
\end{equation}
However, the duality holds for any generic complex background ($N_c$) for which the dual $\theta$ is also complex, see sec. \ref{eq:duality check}. As a consequence, eq. (\ref{eq:quation_of_motion}) remains true for all background/lapse ($N_c$) and hence independent of the metric boundary condition.

%%%%%%%%%%%%%%%%%%%%%%%%%%%%%%%%%%%%%%%%%%%%%%%%

\subsection{Exact path integral and semi-classical limit}
\label{sec:exact-path-integral}
In this appendix, we will briefly sketch the derivation of the exact path integral when the three-form field is coupled to gravity. It is given in eq. (\ref{eq:fixed_flux_path_inte})
\begin{equation}
    \label{eq:fixed_flux_path_inte_appendix}
    \begin{split}
    &Z_n[q_0,q_1]=\int_0^\infty dN_c \int\mathcal{D}q \exp\biggl[iS_{\rm 3-form}(q,N_c,n)\biggr]\\
    \text{where}\quad &S_{\rm 3-form}(q,N_c,n)=2\pi^2\int_0^1 dt\biggl[3N_c-\frac{3\dot{q}^2}{4N_c}-\frac{n^2N_c}{8\pi^4q^2}\biggr].
    \end{split}
\end{equation}
 Introducing the proper time
coordinate $\tau=N_c t$, the action can be written as
\begin{equation}
 S_{\rm 3-form}
 =6\pi^2N_c+
 \int_0^{N_c}d\tau
 \left[
 \frac{m}{2}(q')^2
 -\frac{\hbar^2(\nu^2-\frac14)}{2m q^2}
 \right],
 \quad
 m=-3\pi^2,\quad
 \nu=\sqrt{\frac14-\frac{3n^2}{2\hbar^2}},
 \label{eq:inverse_square_action}
\end{equation}
where $q'=dq/d\tau$. Thus, the $q$-integral is the
radial path integral with an inverse-square potential. Since $\nu^2-1/4<0$, it is an attractive potential. Its exact
kernel is \cite{Grosche:1998yu}
\begin{equation}
 {\cal K}_{\nu}(q_1,q_0;N_c) =
 \frac{3i\pi^2\sqrt{q_0q_1}}{\hbar N_c}
 \exp\!\left[
 -\frac{3i\pi^2(q_0^2+q_1^2)}{2\hbar N_c}
 \right]
 I_\nu\!\left(
 \frac{3i\pi^2q_0q_1}{\hbar N_c}
 \right).
 \label{eq:fixed_lapse_kernel}
\end{equation}
The remaining lapse integral is therefore
\begin{equation}
 Z_n[q_0,q_1]
 =
 \int_0^\infty dN_c\,
 e^{\,i6\pi^2N_c/\hbar}
 {\cal K}_{\nu}(q_1,q_0;N_c).
 \label{eq:remaining_lapse_integral}
\end{equation}
To evaluate the above integral, we utilize the following identities \cite{Ryzhik} (see eq. 6.653.1)
\begin{equation}
    \label{eq:identity_appen}
    \begin{split}
    \int_0^\infty\exp\biggl[-\frac{1}{2}x-\frac{1}{2x}(a^2+b^2)\biggr]I_\nu\biggl(\frac{ab}{x}\biggr)\frac{dx}{x}=& \,2I_\nu(a)K_\nu(b),\quad [0<a<b]\\
    =&\, 2I_\nu(b)K_\nu(a),\quad [0<b<a].
    \end{split}
\end{equation}
Substituting $x=-12i\pi^2N_c/\hbar$, the integral evaluates to (utilizing $\hbar\rightarrow |\hbar| (1-i\ep)$ to make the integral convergent at large $N_c$ )
\begin{equation}
\label{eq:path_integral_appendix}
 Z_n[q_0,q_1]=\frac{3}{i\hbar}\sqrt{q_0q_1}\,
 I_\nu\!\left(\frac{6\pi^2q_<}{\hbar}\right)
 K_\nu\!\left(\frac{6\pi^2q_>}{\hbar}\right).
\end{equation}
$q_<={\rm min}(q_1,q_0),\,\, q_>={\rm max}(q_1,q_0)$
The overall numerical factor can be absorbed into the normalization of the integral measure.

The semiclassical expansion of the exact path integral in the regime away from the Euclidean regime where $\tilde{n}>q_0$ and/or $\tilde{n}>q_1$, can be extracted from the following asymptotic behaviour \cite{DLMF,Ryzhik} 
\begin{equation}
    \label{eq:asymptotic_form_non_euclidean}
    \begin{split}
        &I_{i\mu}(\mu z)\sim \frac{e^{\pi\mu/2}}{\sqrt{2\pi\mu}A}\exp\biggl[-i\mu\Phi(z)-i\pi/4\biggr],\,\,K_{i\mu}(\mu z)\sim \sqrt{\frac{2\pi}{\mu}}\frac{e^{-\pi\mu/2}}{A}\sin\biggl[\mu\Phi(z)+\pi/4\biggr]
    \end{split}
\end{equation}
where, $A=(1-z^2)^{1/4}$, $\Phi(z)=\ln((1+\sqrt{1-z^2})/z)-\sqrt{1-z^2}$ and $0<z<1$. \\

\textbf{Comment on self-adjoint extension:}
It is important to emphasise that the path integral we derive by analytically continuing the radial time-sliced kernel doesn't obey self-adjointness. This is a UV prescription and doesn't change any new exponential structure in the path integral. Any such prescription will replace $I_{\nu}(q_<)$ as a combination of $I_{\nu}(q_<)$ and $K_{\nu}(q_<)$. It will introduce an extra $K_{\nu}(q_<)K_{\nu}(q_>)$ factor in the path integral. The semiclassical exponent corresponding to it is already present in the amplitude. Hence, Such a prescription is therefore irrelevant for the purposes of this paper.

%%%%%%%%%%%%%%%%%%%%%%%%%%%%%%%%%%%%%%%%%%%%%%

\subsection{GS geometry from the Neumann saddles}
\label{app:GS_from_Neumann}
In this appendix, we will explicitly show that the saddles in eq. (\ref{eq:saddle_expression}) associated with the mixed Neumann--Dirichlet condition reproduces the Giddings-Strominger (GS) geometry \cite{Giddings:1987cg}. Let $\mathbf{r}=q(0)$. The
on-shell solution is
\begin{equation}
 q(t)^2=(\mathbf{r}+\alpha N_c t)^2
 -\frac{4\widetilde n^{\,2}N_c^2t^2}{\mathbf{r}^2},
 \label{eq:app_neumann_solution}
\end{equation}
where $\mathbf{r}$ satisfies a quartic equation. At the lapse saddles ($N_{\rho,\eta}$), it simplifies to
\begin{equation}
 \mathbf{r}_\rho=\frac{2\rho\widetilde n}{\sqrt{4+\alpha^2}},
 \qquad
 N_{\rho,\eta}
 =\frac{1}{2}\left[
 \frac{\rho\widetilde n\alpha}{\sqrt{4+\alpha^2}}
 +\eta\sqrt{\widetilde n^2-q_1^2}\right],
 \qquad \rho,\eta=\pm1 .
 \label{eq:app_neumann_saddles}
\end{equation} 
Substituting $\mathbf{r}=\mathbf{r}_\rho$ and $N_c=N_{\rho,\eta}$ into the on-shell 
eq. (\ref{eq:app_neumann_solution}) and taking the derivative, one finds 
\begin{align}
 &\dot q_{\rho,\eta}
 =\frac{N_{\rho,\eta}}{q_{\rho,\eta}}
\left(\alpha \mathbf{r}_\rho-4N_{\rho,\eta}t\right),
 \qquad \text{and}\qquad \left(\alpha \mathbf{r}_\rho-4N_{\rho,\eta}t\right)^2
 +4q_{\rho,\eta}^2
 =(4+\alpha^2)\mathbf{r}_\rho^2=4\tilde{n}^2 .
\end{align}
The above equations ensure that every saddle satisfies the Hamiltonian constraint
\begin{equation}
 \frac{\dot q_{\rho,\eta}^{\,2}}{4N_{\rho,\eta}^2}
 =\frac{\tilde{n}^2}{q_{\rho,\eta}^2}-1 .
 \label{eq:app_Hamiltonian_constraint}
\end{equation}
Setting $q=a^2,\,\dot q=2a\dot a,\, a_{\rm th}^2=\tilde{ n},$
Eq.~\eqref{eq:app_Hamiltonian_constraint} becomes $a^2\dot{a}^2/N_{\rho,\eta}^2=(a_{\rm th}^4-a^4)/a^4$. On each monotonic branch of the saddle, $a$ can be used as the
radial coordinate, so that $dt=da/\dot a$. Hence,
\begin{align}
 -\frac{N_{\rho,\eta}^2}{a^2}dt^2
 =-\frac{N_{\rho,\eta}^2}
 {a^2\dot a^{\,2}}\,da^2
 =\frac{da^2}{1-a_{\rm th}^4/a^4}.
\end{align}
Therefore, the saddle geometry in eq. (\ref{eq:metric}) becomes
\begin{equation}
 ds^2=
 \frac{da^2}{1-\left(a_{\rm th}/a\right)^4}
 +a^2d\Omega_3^2,
 \qquad a_{\rm th}^2=\widetilde n,
 \label{eq:app_GS_metric}
\end{equation}
This is precisely the GS wormhole metric as obtained in eq. (\ref{eq:Giddings-stominger-wormholes}) for the Dirichlet condition on the metric. This establishes that, irrespective of the metric boundary condition, one obtains the same saddle geometry. Thus $\alpha$, $\rho$, and
$\eta$ determines the location, orientation and segment of the
wormhole represented by the saddle, but they drop out of its local
line element. 

To determine which segment of the GS geometry is represented by a given
saddle, let
\begin{equation}
 \alpha=i\ep |\alpha|,\qquad
 \bar{\beta}\equiv\sqrt{4-|\alpha|^2},\qquad
 \Delta_1\equiv\sqrt{q_1^2-\widetilde n^2},
 \qquad 0<|\alpha|<2.
\end{equation}
Substituting
eq.~\eqref{eq:app_neumann_saddles} into
eq.~\eqref{eq:app_neumann_solution}, the solution can be written as
\begin{equation}
 q_{\rho,\eta}(t)^2
 =\tilde {n}^2+\Xi_{\rho,\eta}(t)^2,
 \qquad
 \Xi_{\rho,\eta}(t)=\left(
 \frac{\rho\ep\tilde{n}|\alpha|}{\bar{\beta}}
 +\eta\Delta_1
 \right)t-\frac{\rho\ep\tilde{n}|\alpha|}{\bar{\beta}}.
 \label{eq:app_signed_GS_coordinate}
\end{equation}
Here $\Xi$ is a signed radial coordinate on the GS wormhole:
$\Xi=0$ is the throat, while the two signs of $\Xi$ distinguish the
two sides of the throat. Its overall sign is conventional, but the relative sign of the two boundaries is physical. At the
boundaries,
\begin{equation}
\label{eq:signed_radial_coordinate}
 \Xi_{\rho,\eta}(0)
 =-\frac{\rho\ep \tilde{n}|\alpha|}{\bar{\beta}},
 \qquad
 \Xi_{\rho,\eta}(1)=\eta\Delta_1,
 \qquad
 \Xi_{\rho,\eta}(0)\Xi_{\rho,\eta}(1)
 =-\rho\ep\eta\,
 \frac{\tilde{n} |\alpha|\Delta_1}{\bar{\beta}}.
\end{equation}
Consequently,
\begin{equation}
\label{eq:saddle_classfication}
 \rho\ep\eta=+1
 \ \Longrightarrow\ \text{cross-throat},
 \qquad
 \rho\ep\eta=-1
 \ \Longrightarrow\ \text{same-side}.
 \;
\end{equation}
Indeed, on the positive-$\mathbf{r}_\rho$ sheet with $\rho=+1$ and
$\alpha=-i|\alpha|$ ($\ep=-1$), the $\eta=-1$ saddle has
$\Xi(0)>0$ and $\Xi(1)<0$ and therefore crosses the throat, whereas the
$\eta=+1$ saddle has $\Xi(0)>0$ and $\Xi(1)>0$ and remains on the same
side. At $|\alpha|=0$, the initial boundary lies at $\Xi(0)=0$, giving the half-wormhole geometry. The cases $|\alpha|\rightarrow 2$ for the cross-throat branch correspond to a full imaginary wormhole. A generic complex $\alpha$ instead yields the corresponding complexified GS geometry. 

%%%%%%%%%%%%%%%%%%%%%%%%%%%%%%%%%%%%%%%%%%%%%%

\subsection{Singular histories and the physical lift of Lorentzian contour}
\label{sec:physical_lift}
In this appendix, we discuss about the geometries which are singular, i.e., $q(t_*)=0$ for $t_*\in (0,1)$. Consider the covering coordinate as mentioned in eq. (\ref{eq:covering_map_neumann}). Substituting $u=+i\ep \pi/2+u_x$ and $\alpha=i\ep|\alpha|$ in $N_c(u)$ and $\mathbf{r}(u)$, we get
\begin{equation}
    \label{eq:covering_map_horizontal_line}
    N_c(u_x)=\frac{q_1^2\cosh(u_x)\sinh(u_x)}{|\alpha|q_1\cosh(u_x)-2\tilde{n}},\quad \mathbf{r}(u_x)=i\frac{2\tilde{n}q_1\sinh(u_x)}{|\alpha|q_1\cosh(u_x)-2\tilde{n}}\equiv i\mathbf{R}
\end{equation}
When the denominator is positive $u_x>0$ maps to $N_c>0$ with $\mathbf{R}>0$. Computing $q(t)$ along the line, we obtain
\begin{equation}
    \label{eq:singular_history}
    q(t_*)=0,\quad \text{where,} \quad t_*=\frac{\mathbf{R}}{\mathbf{R}+q_1 e^{u_x}},\quad t_*\in(0,1).
\end{equation}
Now, the problem arises if there is a ramification point on this horizontal line connecting the physical lift to the singular histories. The equation for the ramification point is $dN_c/du=0$, along the horizontal line, which reduces to
\begin{equation}
    \label{eq:ramification_horizontal_line}
    -2\tilde{n}\cosh^2(x)+|\alpha| q_1 \cosh^3(x)-2\tilde{n}\sinh^2(x)=0.
\end{equation}
The above equation has a real $x$ solution if and only if
\begin{equation}
    \label{eq:ramification_condition}
    \beta\equiv \frac{|\alpha|q_1}{2\tilde{n}},\quad \text{such that}\quad \beta <\beta_{\rm crit},\quad \text{where,} \quad \beta_{\rm crit}=4\sqrt{6}/9.
\end{equation}
This is easy to see by writing eq. (\ref{eq:ramification_horizontal_line}) as $\beta=2/c-1/c^3$, $c\equiv\cosh(x)$ and observing that the function in the right has a maximum when $c=\sqrt{3/2}$. At this value $\beta$ is $4\sqrt{6}/9$.
When $\beta<\beta_{\rm crit}$, the physical lift starting at $u=0$ at $N_c=0$ hits the horizontal line of singular histories. To avoid this, we rotate the contour $N_c=\rho e^{i\delta},0<\delta\ll 1$, as shown in fig. \ref{fig:physical_lift_rotating contour}. Then $N_c(u)\sim R_\ep+i I_\ep$. The equation of the lifted contour becomes $\arg(N_c)=\delta,\rho>0$,
\begin{equation}
    \label{eq:lifted_contour_equation}
   \delta R_\ep-I_\ep=0,\qquad R_\ep+\delta I_\ep >0,
\end{equation}
where $R_\ep$ and $I_\ep$ are defined as 
\begin{equation}
\label{R_and_I}
    \begin{split}
&R_\ep=2\tilde{n}\sinh(u_x)\cosh(u_x)\cos(2u_y)+\ep|\alpha|q_1\sinh(u_x)\sin(u_y)(\sinh^2 u_x+\sin^2u_y)\\
        &I_\ep=\cos(u_y)[2\tilde{n}\cosh(2u_x)\sin u_y-\epsilon |\alpha| q_1\cosh u_x(\sinh^2 u_x+\sin^2u_y)].
    \end{split}
\end{equation}

%%%%%%%%%%%%%%%%%%%%%%%%%%%%%%%%%%%%%%%%%%%%%
\subsection{Comment on the scaling symmetry}
\label{sec:symmetry}
In this appendix, we examine the scaling transformations of the metric and the three-form field and determine how the action, boundary conditions, and Picard--Lefschetz intersection numbers transform under these rescalings. This is a consequence of considering a flat-space wormhole. Under the scaling of the metric $g_{\mu\nu}\rightarrow \lam g_{\mu\nu}$ as in eq. (\ref{eq:metric}), one gets
\begin{equation}
    \label{eq:metric_scaling}
    g_{\mu\nu}\rightarrow \lam g_{\mu\nu}\quad \Rightarrow N_c\rightarrow \lam N_c, \quad \text{and} \quad q\rightarrow \lam q.
\end{equation}
Under the above rescaling, the Dirichlet condition (fixes $q$) rescales while the Neumann condition ($\dot{q}/N_c$) remains invariant. As a consequence of this, one requires no background subtraction of divergence for the Neumann end of the boundary condition. This also enables us to sum over $n$, for $\alpha$, from its value away from the imaginary critical value. This in turn yields a bound relating the imaginary part of the boundary axion and $\alpha$. 

The metric rescaling affects the relevance of saddles. For fixed $\tilde{n}$, the relevance of saddles can change depending on whether one has any of the three regions \cite{Loges:2022nuw} i) $q_0<q_1<\tilde{n}$, ii) $q_0<\tilde{n}<q_1$ and iii) $\tilde{n}<q_0<q_1$. Keeping fixed $n$, if one rescales $q_0\rightarrow\lam q_0,q_1\rightarrow\lam q_1$, the situation in i) and ii) modifies to iii). This modifies the relevant saddles. However, when we are in the Euclidean regime (iii), such rescaling has no effect. In this regime, it is legitimate to rescale the geometry without changing the intersection number.

The action for the three-form scalar side of the duality takes the form
\begin{equation}
    \label{eq:three-form-appen}
    S_{\rm 3-form}(q,N_c,n)=2\pi^2\int_0^1 dt\biggl[\underbrace{3N_c-\frac{3\dot{q}^2}{4N_c}}_{S_{\rm EH}}-\underbrace{\frac{n^2N_c}{8\pi^4q^2}}_{S_n}\biggr]
\end{equation}
Under the metric transformation in eq. (\ref{eq:metric_scaling}) $S_{\rm EH}\rightarrow\lam S_{\rm EH}$ and $S_n\rightarrow S_n/\lam$. This opposite scaling is the reason behind the presence of a wormhole \cite{Witten:2026twr}. Similar scaling behaviour is seen for the action in eq. (\ref{eq:3-form-neumann}) Neumann boundary condition. Moving to the axion side of the duality, we note that the action in eq. (\ref{eq:fixed_charge_path_int}) obeys the following scaling
\begin{equation}
    \label{eq:scaling_axion_appen}
    S_{\rm axion}[\theta,\lam q,\lam N_c]= \lam S_{\rm axion}[\theta,q,N_c].
\end{equation}
Under the scaling, the Dirichlet saddles in eq. (\ref{sec:saddle_dirichlet}) transform as $N_{\ep_1\ep_2}\rightarrow\lam N_{\ep_1\ep_2}$. However, the Neumann saddles in eq. (\ref{eq:action_neuman_Scalar}) remains unchanged. To make the duality in eq. (\ref{eq:stress-tensor_equality}) same one also require $\tilde{n}$ to scale as $\tilde{n}\rightarrow\lam\tilde{n}$. Performing the scaling, we obtain
\begin{equation}
    \label{eq:scaling_axion}
    S_{\rm 3-form}(\lam q,\lam N_c,\lam \tilde{n})= \lam S_{\rm 3-form}(q,N_c,\tilde{n}).
\end{equation}
Such scaling has no effect on the PL analysis.

%%%%%%%%%%%%%%%%%%%%%%%%%%%%%%%%%%%%%%%%%%%%%

\subsection{An inequality for the KSW non-allowable region}
\label{sec:KSW_allowability_append}
\begin{figure}
    \centering
    \includegraphics[width=0.65\linewidth]{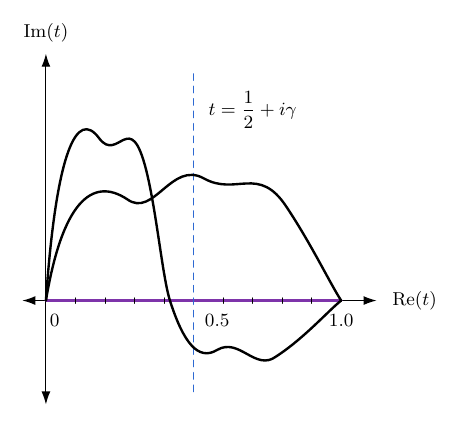}
    \caption{The black curves are the examples of complex time paths starting and ending at $t=0$ and $t=1$, respectively, which always intersect the vertical line at $t=1/2+i\gamma,\,\,\gamma\in\mathbb{R}$.}
    \label{fig:time_path}
\end{figure}
In this appendix, we will derive the eq. \ref{eq:KSW_allowability_u_plane}. As shown in fig. \ref{fig:time_path}, any continuous complex time path always intersects $t=1/2+i\gamma,\gamma\in\mathbb{R}$. Hence, it is sufficient to analyze the allowability for points on this line. To start with, substituting $t=1/2+i\gamma,\gamma\in\mathbb{R}$ in $\mathbf{Q}_\gamma(u)=q^2(t;u)$, we get

\begin{equation}
    \label{eq:q(t)_rho}
    \frac{\mathbf{Q}_\gamma(u)}{2q_0q_1}=\frac{m+\cosh(u)}{4}+\gamma^2(\cosh(u)-m)+i\gamma d,
\end{equation}
where $m=(q_0^2+q_1^2)/2q_0q_1, d=(q_1^2-q_0^2)/2q_0q_1$ and $m^2-d^2=1, m\geq 1$. Then, according to eq. (\ref{eq:ksw_non-allowable_argument}), the geometry $u$-will be KSW non-allowable if
\begin{equation}
    \label{eq:Q_arg_appen}
    |\arg (\mathbf{Q}_\gamma(u))|\geq 2\pi/3,\quad \forall \gamma\in\mathbb{R}.
\end{equation}
On writing $\mathbf{Q}_\gamma=R_\gamma+iI_\gamma$, the above inequality is equivalent to the following set of conditions 
\begin{equation}
    \label{eq:h_sigma}
    h_\sigma(\gamma)\equiv \sqrt{3}R_\gamma+\sigma I_\gamma\leq 0,\quad \sigma=\pm 1.
\end{equation}
On writing $\cosh(u)=c_R+i c_I$, $h_\sigma(\gamma)$ becomes the quadratic polynomial of $\gamma$
\begin{equation}
    \label{eq:quadratic_polynomial}
    \begin{split}
&h_\sigma(\gamma)=A_\sigma\gamma^2+B_\sigma\gamma+C_\sigma\\
\text{where,}\quad & A_\sigma=\sqrt{3}(c_R-m)+\sigma c_I,\quad B_\sigma=\sigma d,\quad C_\sigma=\frac{\sqrt{3}(m+c_R)+\sigma c_I}{4}.
\end{split}
\end{equation}
Now, $h_\sigma(\gamma)$ will be negative for all $\gamma\in\mathbb{R}$, if $A_\sigma<0$ and discriminant $\Delta_\sigma=4 m^2-1-(\sqrt{3 }c_R+\sigma c_I)^2<0$. This implies the following inequality
\begin{equation}
    \label{eq:inequality_ksw}
    \sqrt{3}c_R+\sigma c_I\leq -\sqrt{4 m^2-1}\quad \forall \sigma=\pm 1
\end{equation}
where we have utilized that for $m\geq 1$, $\sqrt{4 m^2-1}\geq \sqrt{3} m$. Since the inequality in eq. (\ref{eq:inequality_ksw}), has to be valid for both the signs of $\sigma$, we have $\max(\sqrt{3}c_R+ c_I,\sqrt{3}c_R-c_I)\leq -\sqrt{4 m^2-1}$. It further implies $\sqrt{3}c_R+ |c_I|\leq -\sqrt{4 m^2-1}$. Now
\begin{equation}
    \label{eq:cosh(u)}
\cosh(u_x+iu_y)=\cosh(u_x)\cos(u_y)+i\sinh(u_x)\sin(u_y)=c_R+ic_I.
\end{equation}
Hence, we obtain the following inequality, as in eq. (\ref{eq:KSW_allowability_u_plane})
\begin{equation}
    \label{eq:KSW_allowability_u_plane_appen}
    \sqrt{3}\cosh(u_x)\cos(u_y)+ |\sinh(u_x)\sin(u_y)|\leq -\sqrt{4 m^2-1}.
\end{equation}
As explained in the main text, if the above inequality is satisfied, then there is no continuous complex-time path connecting $t=0$ and $t=1$ that can make the geometry in the $u$-plane KSW-allowed.

%%%%%%%%%%%%%%%%%%%%%%%%%%%%%%%%%%%%%%%%%%%%%%%%%%%%%%%%

\end{document}